\documentclass[10pt]{article}

\usepackage{graphicx}%
\usepackage{amsmath,amssymb,amsthm,eucal,upref,bm}%
\usepackage{labelfig}%

\begin{document}

\baselineskip=4.6mm

\makeatletter

\newcommand{\E}{\mathrm{e}\kern0.2pt}
\newcommand{\D}{\mathrm{d}\kern0.2pt}
\newcommand{\RR}{\mathrm{I\kern-0.20emR}}

\def\bottomfraction{0.9}

\title{{\bf Dispersion equation for water waves \\ with vorticity and Stokes waves
\\ on flows with counter-currents}}

\author{Vladimir Kozlov$^a$, Nikolay Kuznetsov$^b$}

\date{}

\maketitle

\vspace{-8mm}

\begin{center}
$^a${\it Department of Mathematics, Link\"oping University, S--581 83 Link\"oping,
Sweden \\ $^b$ Laboratory for Mathematical Modelling of Wave Phenomena, \\ Institute
for Problems in Mechanical Engineering, Russian Academy of Sciences, \\ V.O.,
Bol'shoy pr. 61, 199178 St Petersburg, Russian Federation} \\

\vspace{2mm} 

E-mail: vlkoz@mai.liu.se / V.~Kozlov; nikolay.g.kuznetsov@gmail.com / \\ N.~Kuznetsov
(corresponding author, tel.: 7 (812) 4334873, fax: 7 (812) 3214771)
\end{center}

\begin{abstract}
The two-dimensional free-boundary problem of steady periodic waves with vorticity is
considered for water of finite depth. We investigate how flows with small-amplitude
Stokes waves on the free surface bifurcate from a horizontal parallel shear flow in
which counter-currents may be present. Two bifurcation mechanisms are described: for
waves with fixed Bernoulli's constant and fixed wavelength. In both cases the
corresponding dispersion equations serve for defining wavelengths from which Stokes
waves bifurcate. Necessary and sufficient conditions for the existence of roots of
these equations are obtained. Two particular vorticity distributions are considered
in order to illustrate general results.

\vspace{2mm}


\end{abstract}

\section{Introduction}

\setcounter{equation}{0}

We study the two-dimensional nonlinear problem of steady waves in a horizontal open
channel of uniform rectangular cross-section occupied by an inviscid, incompressible
and heavy fluid, say, water. The water motion is assumed to be rotational which,
according to observations, is the type of motion commonly occurring in nature (see,
for example, \cite{SCJ,Th} and references cited therein). Our aim is to consider the
bifurcation mechanisms resulting in the formation of Stokes waves (periodic waves
whose profiles rise and fall exactly once per wavelength) on a horizontal free
surface of a parallel shear flow in which counter-currents may be present. One of
these mechanisms keeps Bernoulli's constant fixed which is convenient for
approximating solitary waves. On the other hand, keeping the wavelength fixed as in
the second approach is convenient for obtaining global branches of solutions and
investigating waves of extreme height. 

A detailed study of shear flows, which forms the basis for present approaches, is
given in our previous paper \cite{KK} under the assumption that the flow velocity
depends only on the vertical coordinate. The crucial point is to determine the set
of bifurcation wavelengths for which purpose dispersion equations are used. This
technique is similar to the method that was earlier applied in our investigation of
the irrotational Stokes waves (see \cite{KK1}, section~8.2.1). To the authors' best
knowledge, the dispersion equations derived here (see (\ref{dispers}) in section~1.4
and (\ref{dispers_*}) in section~1.5) were not used previously. In order to
illustrate how the general results work for particular vorticity distributions
several examples are presented in section~6.

\subsection{Statement of the problem}

Let a horizontal open channel of uniform rectangular cross-section be bounded below
by the horizontal rigid bottom. Let water occupying the channel be bounded above by
a free surface. The surface tension is neglected and the pressure is assumed to be
constant on the free surface. The assumption that the water motion is
two-dimensional and rotational and the incompressibility of water allows us to seek
the velocity field in the form $(\Psi_Y, -\Psi_X)$, where $(X,Y)$ are appropriate
Cartesian coordinates and $\Psi$ is a {\it stream function} (see, for example,
\cite{LSh} and \cite{S} for its definition). The vorticity distribution $\omega
(\Psi)$ is supposed to be a prescribed $C^{2,\alpha}_{loc}$-function on $\RR$,
$\alpha \in (0,1)$, with bounded derivative. In the present paper, all variables are
non-dimensional and chosen so that the constant volume rate of flow per unit span
and the constant acceleration due to gravity are scaled to unity in the same way as
in the classical paper \cite{KN} by Keady and Norbury. Namely, lengths and
velocities are scaled to $(Q^2/g)^{1/3}$ and $(Qg)^{1/3}$, respectively; here $Q$
and $g$ are the dimensional quantities for the volume rate of flow per unit span and
the gravity acceleration, respectively (see Appendix~A for the details of scaling).
We recall that $(Q^2/g)^{1/3}$ is the depth of the {\it critical uniform} stream in
the irrotational case (see, for example, \cite{Ben}).

Let $(X,Y)$ be such that the $X$-axis lies in the longitudinal section of the
canal's bottom and gravity acts in the negative $Y$-direction. Moreover, let the
frame of reference be chosen so that the velocity field and the unknown free surface
are time-independent in these coordinates. Assuming that $Y = \xi (X)$ represents
the free-surface profile (here $\xi$ is a positive, continuous function), we denote
by ${\cal D}$ the longitudinal section of the water domain, that is,
\[ {\cal D} = \{ -\infty < X < +\infty ,\ 0 < Y < \xi (X) \} .
\]

Since the surface tension is neglected, $\Psi$ and $\xi$ must satisfy the following
free-boundary problem:
\begin{eqnarray}
&& \Psi_{XX} + \Psi_{YY} + \omega (\Psi) = 0,\quad (X,Y) \in {\cal D} ; \label{1} \\
&& \Psi (X, 0) = 0,\quad X\in \RR; \label{2} \\ && \Psi (X, \xi (X)) = 1, \quad X\in
\RR; \label{3} \\ && |\nabla_{X,Y} \Psi (X, \xi (X))|^2 + 2 \xi (X) = 3 r , \quad
X\in \RR . \label{4}
\end{eqnarray}
In the last relation (Bernoulli's equation), $r > 0$ is the problem's parameter
referred to as the non-dimensional total head (also known as Bernoulli's constant).
This statement for steady irrotational waves was proposed by Benjamin and Lighthill
in their groundbreaking paper \cite{BL}, where the renowned conjecture was
formulated (see also \cite{Ben} and \cite{KK1,KK0}). The problem with $\omega \neq
0$ formulated above appeared in \cite{KN}. It has various advantages, the most
significant of which is its equivalence to Hamiltonian systems based on the flow
force invariant as Hamiltonian. This equivalence was established for waves with
vorticity in the recent paper \cite{KK5}, where references concerning the
irrotational case can be found. Furthermore, some effects distinguishing waves with
vorticity from irrotational ones have been discovered on the basis of this
formulation. In particular, the existence of local minima on the
wavelength-streamdepth curves was found numerically in \cite{Do}, and the fact that
no steady water waves of small amplitude are supported by a shear flow with a still
free surface was established rigorously in \cite{KK4}.

The free-boundary problem describing water waves with vorticity has long been known
(see details in section 1.2 below); its derivation from the governing equations and
the assumptions about the boundary behaviour of water particles can be found in
\cite{CS}. Reformulation of the problem based on the partial hodograph transform was
proposed in \cite{DJ}, but this transform is possible only for unidirectional flows.
Other reformulations are given in sections~2 and 5.

Considering the question of formation of Stokes waves on a flat free surface of a
shear flow, we suppose that $\Psi \in C^{2,\alpha} (\bar{\cal D})$ and $\xi \in
C^{2,\alpha}$, $0<\alpha<1$. These assumptions can be relaxed using a weak
formulation, in which case $\Psi$ and $\xi$ occur to be $C^{1,\alpha}$-functions
(see, for example, \cite{VZ}).

\subsection{Background}

Nonlinear theory of gravity water waves with vorticity has a long history which
dates back to the 1800s, when Gerstner \cite{G} found his remarkable explicit
solution (see \cite{C} for a modern approach to this solution; in \cite{Hen} some
its properties are obtained). Note that Scott Russell and Stokes published their
pioneering works about irrotational water waves only in the 1840s (see \cite{SR} and
\cite{St}, respectively). Nevertheless, rotational waves have been studied to a less
extent than irrotational ones (see the recent survey paper \cite{Str}, where an
extensive bibliography is provided). Only a few articles treating waves with
vorticity rigorously were published during the 20th century, and at least three of
them are of lasting interest. As early as 1934, Dubreil-Jacotin \cite{DJ} had proved
the first existence result for these waves (in the 1950s, her work was extended by
Goyon \cite{Goy}). For this purpose she introduced a partial hodograph transform
which is a very convenient tool to investigat rotational waves provided the
horizontal component of the relative velocity in the flow does not change sign. In
1978, Keady and Norbury \cite{KN} obtained bounds on the total head of flow and
free-surface profiles, but their results were proved only for a rather small class
of vorticity distributions. In the authors' work \cite{KK3}, these bounds were
generalized to the case of arbitrary Lipschitz distributions.

During the past decade, a substantial body of rigorous results about waves with
vorticity has appeared, the first of which was the article \cite{CS} by Constantin
and Strauss. They used the partial hodograph transform for obtaining a global branch
of large-amplitude, Stokes waves. No counter-currents are allowed in \cite{CS} as
well as in the most of other papers that are briefly characterized by placing them
into the following overlapping groups each of which covers a particular topic.

\vspace{2mm}

\noindent $\bullet$ The existence of waves is shown through the local/global
bifurcation mechanism in \cite{BT,CS4,CV,GW,Hen2,H0,H1,H2,V1,V2,W-1,W0,W}.

\noindent $\bullet$ Regularity (in particular, analyticity) of the stream function
and/or streamlines is investigated in \cite{CE1,CV,Esch,H4,M,MM}.

\noindent $\bullet$ Symmetry of periodic and solitary waves is studied in
\cite{CEW,CE0,CE,E,H,H3,MM}.

\noindent $\bullet$ Stability and instability of waves is considered in \cite{CS3}
and \cite{HL}, respectively.

\noindent $\bullet$ The works \cite{C1,CSS,CS2,CS4,Do,E-1,E0,E1,E2,EV,K,V,V1,V2,VZ,W,WZ}
deal with some other properties of waves.

\noindent $\bullet$ Deep-water waves are treated in \cite{CE0,E,H0,H2}.

\noindent $\bullet$ Solitary waves are investigated in 
\cite{GW,Hen1,H1,H3,MM,Ter,Wheel1}.

\noindent $\bullet$ The papers \cite{C3,C2,CIP,EEV,GW,KK5} are devoted to various forms
of a dynamical system approach to waves with vorticity.

\noindent $\bullet$ Numerical results about rotational waves are presented, for
example, in \cite{Do,F,KS,KS1,Per,VB1,VB2,VB3}.

\noindent $\bullet$ Dispersion relations are considered in \cite{C4,CS4,Hen3,KKL}
for periodic traveling waves; the first three of these papers deal with
unidirectional flows and discontinuous vorticity distributions.

\vspace{2mm}

Now we turn to a few articles \cite{CV,EEV,EEW,EV,Wah}, in which rotational waves on
flows with counter-currents are studied (the last of these papers contains an
extensive list of references and a brief review of the literature on water waves
with vorticity). The works \cite{CV,EV,Wah} are concerned with the case of a
constant vorticity, whereas a linear vorticity is considered in \cite{EEV,EEW}.
Since the results obtained in the cited papers are closely related to ours, we
discuss them in greater detail.

It should be emphasized that counter-currents are not considered as the principal
point of those articles. Instead, the authors concentrate their attention on the
appearance of internal {\it stagnation points} and the so-called {\it critical
layers}. A critical layer is a connected subset of the water domain consisting of
closed streamlines and stagnation points. See, for example, figures~1 and 3 in
\cite{Wah} and figure~3 in \cite{CV}. However, there exist degenerate cases like
that shown in figure~2, \cite{Wah}. From a geometric point of view critical layer is
a horizontal array of cat's eye patterns anticipated by Kelvin \cite{Thom} (see also
\cite{DR}, p.~141). Therefore, such a layer, generally speaking, separates two other
layers with opposite directions of flow.

First, the existence of steady waves on a flow of constant vorticity in which a
critical layer (and so, a counter-current) is present was established by Ehrnstr\"om
and Villari \cite{EV}, who studied streamlines and particle paths in the framework
of linear theory. It was Wahl\'en \cite{Wah}, who proved the existence of
small-amplitude Stokes waves with a constant vorticity and discovered that in the
reference frame moving with the wave there is a critical layer. (As is discussed in
\cite{Wah}, the term itself has a long history, but was mainly used in the framework
of mathematical models other than rotational water waves.) Recently, Constantin and
Varvaruca \cite{CV} proposed an approach to Stokes waves with a constant vorticity
that differs from that in \cite{Wah} and is based on the conformal mapping technique
that imposes no restriction on the geometry of the free surface profile which, in
particular, can be overhanging. This technique, as the authors claim, opens up the
way to using global bifurcation theory. Besides, the physical relevance of the
problem with a constant vorticity is discussed in \cite{CV}.

In the articles \cite{EEV,EEW}, Ehrnstr\"om {\it et al.} studied the case of linear
vorticity when multiple counter-currents and critical layers exist. In \cite{EEW},
the results obtained in \cite{Wah} were extended, yielding the existence of
small-amplitude waves with arbitrarily many critical layers (including the so-called
bichromatic waves). Qualitative and some quantitative properties of waves the
existence of which is proved in \cite{EEW} are investigated in \cite{EEV}. In
particular, a classification of vorticity distributions is proposed and the
bifurcation relations are given for different cases.

\subsection{Stream solutions}

A pair $(u (Y), \, h)$, where $h = {\rm const}$, is called a {\it stream solution}\,
of problem (\ref{1})--(\ref{4}) when $\Psi (X,Y) = u (Y)$ and $\xi (X) = h$ satisfy
this problem. Such solutions are studied in detail in \cite{KK}, where they are also
discussed in terms of the unified theory of conjugate flows developed by Benjamin
\cite{Benuni}. A summary of results obtained in \cite{KK} is given in Appendix B and
here we restrict ourselves to notions required in formulations of our main theorems.

Seeking stream solutions, we write problem (\ref{1})--(\ref{4}) in the form:
\begin{equation}
u'' + \omega (u) = 0 \ \mbox{on} \ (0, h) , \ \ u (0) = 0, \ \ u (h) = 1, \ \ |u'
(h)|^2 +2 h = 3 r , \label{ss1}
\end{equation}
where $u' = u_Y$. By $U (Y; s)$, we denote a unique solution existing on the whole
$\RR$ for the first equation (\ref{ss1}) complemented by the following Cauchy data:
\begin{equation}
U (0; s) = 0,\quad U' (0; s) = s , \quad {\rm where}\ s \in \RR . 
\label{cauchy}
\end{equation}
It occurs that all stream solutions are parameterised by
\begin{equation}
s \geq s_0 = \sqrt{2 \max_{0 \leq \tau \leq 1} \Omega (\tau)} \ , \quad 
\mbox{where}\ \Omega (\tau) = \int_0^\tau \omega (t) \, {\D} t . \label{s_0}
\end{equation}

Now we summarize our algorithm that gives stream solutions from which Stokes waves
can bifurcate for a given $r$. First, appropriate values of $s$ must be determined
from Bernoulli's equation (see the second relation (\ref{ber_tr}) in Appendix~B):
\begin{equation}
{\cal R}_j^{(\pm)} (s) = r , \quad {\rm where} \ {\cal R}_j^{(\pm)} (s) =
\frac{1}{3} \left[ s^2 - 2 \, \Omega (1) + 2 \, h_j^{(\pm)} (s) \right] , \quad j =
0,1,\dots  \label{ss2}
\end{equation}
Here the values of depth $h_j^{(\pm)} (s)$ are given by formulae
(\ref{h2k})--(\ref{hj-}) (see Appendix~B); their number (finite or infinite) depends
on the vorticity distribution. Then the stream solution corresponding to the root
$s$ of equation (\ref{ss2}) with the subscript $j$ and the superscript $(+)$ $[(-)]$
is as follows: 
\[ \left( U (Y; s) , h_j^{(+)} (s) \right) \quad \left[ \left( U (Y; -s) , h_j^{(-)} 
(s) \right) \right] .
\]
Thus, either of the latter pairs gives $u (Y)$ and $h$ satisfying problem
(\ref{ss1}).

Since for every $s > s_0$ the sequences ${\cal R}_j^{(+)} (s)$ and ${\cal R}_j^{(-)}
(s)$ are increasing (see Appendix~B), we have that the graph of ${\cal R}_0^{(+)}
(s)$ (it is convex as is shown in \cite{KK}, section~5.1) lies below those
corresponding to other functions ${\cal R}_j^{(\pm)} (s)$. Therefore, we put
\begin{equation}
r_c = \min_{s \geq s_0} {\cal R}_0^{(+)} (s) \label{ss3}
\end{equation}
and this value is always attained at some $s_c > s_0$. Moreover, for all $r > r_c$
some equations in the sequence (\ref{ss2}) have roots; at least two such roots exist
for every $r$. Thus, for each $s$ obtained from the sequence of equations
(\ref{ss2}) one finds $U (Y; s)$ and $h (s)$, and so the set of all stream solutions
corresponding to $r \geq r_c$ is described.

\subsection{Bifurcation with fixed Bernoulli's constant: \\ formulation of
main results}

When Bernoulli's constant is fixed, the existence theorem for Stokes waves involves
two assumptions. The first assumption is aimed at determining a shear flow from
which Stokes waves bifurcate, namely, we suppose the following.

\vspace{2mm}

\noindent (I) The inequality $r > r_c$ holds and there exists a stream solution $(U,
h)$ corresponding to $r$ such that $U' (h) \neq 0$.

\vspace{2mm}

As was said above, some equations in the sequence (\ref{ss2}) have at least two
roots roots when $r > r_c$ (see details in \cite{KK}). Moreover, one of them, say,
$s_*$ defines the stream solution for which $U' (h; s_*) \neq 0$ (it is explained in
the first paragraph of section~1.5 that $s_* > s_0$ unless $s_0 = 0$).

The next assumption concerns the so-called dispersion equation and its roots.
Similarly to the case of irrotational waves, each of them determines a bifurcation
wavelength as a function of $r$ and $h$ (of course, $h$ itself depends on $r$). In
the case of waves with vorticity, more than one bifurcation wavelength can exist for
a given choice of a stream solution and this depends on the vorticity distribution
(see section 6). On the contrary, only one wavelength exists for a given admissible
$r$ in the irrotational case because the corresponding dispersion equation can be
written as follows (see \cite{KK1}, p.~478):
\begin{equation}
\tau \coth h \tau - h^2 = 0 . \label{coth}
\end{equation}
Here $h > 1$ is defined by $r$ and its value is equal to the non-dimensional depth
of the subcritical uniform stream (for convenience, the depth of the critical stream
is scaled to unity). It is clear that (\ref{coth}) has only one positive root which
is equal to the wavenumber $2\, \pi / \Lambda_0$ corresponding to the bifurcation
wavelength.

\vspace{2mm}

In the latter case, the stream solution $(U (\cdot; s_*), \, h (s_*))$ corresponding
to the chosen $s_*$ yields the following dispersion equation (see its derivation in
section~2.3):
\begin{equation}
\sigma (\tau) = 0 ,  \quad \mbox{where} \ \ \sigma (\tau) = \kappa \, \gamma' (h,
\tau) - \kappa^{-1} + \omega (1) \ \ \mbox{and} \ \ \kappa = U' (h) ,
\label{dispers}
\end{equation}
whereas $\gamma (Y, \tau)$ solves the following problem:
\begin{equation}
- \gamma'' + [ \tau^2 - \omega' (U) ] \, \gamma = 0 \ \mbox{on} \ (0, h), \quad
\gamma (0, \tau) = 0 , \ \ \gamma (h, \tau) = 1 . \label{gamma}
\end{equation}
Here $h$ depends on $r$ through the root $s_*$ of Bernoulli's equation. Note also
that
\[ \kappa = U' (h) = \pm \sqrt{3r - 2h} \quad \mbox{when} \ \pm \, U' (h) > 0 .
\]
If $\tau^2$ is not a Dirichlet eigenvalue of the operator $\D^2 / \D Y^2 + \omega'
(U)$ (there might exist a finite number of such eigenvalues), then problem
(\ref{gamma}) has a unique solution. Therefore, $\gamma (Y, \tau)$ and $\sigma
(\tau)$ are defined for all these values of $\tau$ and are smooth and even functions
of $\tau$. However, there are vorticity distributions for which the Dirichlet
spectrum is not empty, and so $\gamma (Y, \tau)$ and $\sigma (\tau)$ are not defined
for some values of $\tau$. For example, this is true when $\omega (\tau)$ is a
linear function positive for $\tau > 0$ (see section~6.2). Therefore, we will
consider (\ref{dispers}) under the assumption that $\tau^2$ is not a Dirichlet
eigenvalue of $\D^2 / \D Y^2 + \omega' (U)$.

\vspace{2mm}

There is an alternative to the dispersion equation (\ref{dispers}), namely, to
determine $\tau^2$ from the following problem of Sturm--Liouville type:
\[ - G'' + [ \tau^2 - \omega' (U) ] \, G = 0 \ \mbox{on} \ (0, h), \quad
G (0) = 0 , \ \ \kappa \, G' (h) = \left[ \kappa^{-1} - \omega (1) \right] G (h) .
\]
Indeed, this problem has a non-trivial solution corresponding to $\tau^2_0$ if and
only if $\tau_0$ satisfies equation (\ref{dispers}). A similar approach was applied,
in particular, by Constantin and Strauss in \cite{CS,CS4}) for unidirectional flows.

The advantage of (\ref{dispers}) is threefold. First, this equation generalizes
(\ref{coth}) arising in the irrotational case. Indeed, if $\omega$ vanishes
identically, then (\ref{dispers}) coincides with (\ref{coth}) because $\kappa =
h^{-1}$ and $U = Y / h$, and so the solution of (\ref{gamma}) is $\gamma = \sinh Y
\tau / \sinh h \tau$ which gives (\ref{coth}). Second, (\ref{dispers}) has another
property common with (\ref{coth}) (see Lemma 1.1 below and the comment preceding
it). What is most important, the transversality condition is expressed in terms of
the function $\sigma$ when one applies the Crandall--Rabinowitz theorem for proving
the existence of local bifurcation (see Propositions 4.2 and 5.1).

\vspace{2mm}

Let us list some properties of equation (\ref{dispers}). First, the function $\sigma
(\tau)$ is well-defined provided $\kappa \neq 0$ (this is true when $s_* \neq s_0 >
0$). Second, if zero is not a Dirichlet eigenvalue of $\D^2 / \D Y^2 + \omega' (U)$
(this is equivalent to the inequality $(\D h /\D s) \, (s_*) \neq 0$; see
Remark~3.4), then the following equality holds (see formula (\ref{sigma(0)>}) in
Proposition~3.3):
\[ \sigma (0) = - \frac{3}{2 \kappa} \left[ \frac{\D {\cal R}}{\D s} (s_*) 
\bigg/ \frac{\D h}{\D s} (s_*) \right] .
\]
For the sake of brevity, here and below ${\cal R}$ stands for the function ${\cal
R}_j^{(\pm)}$ that defines the stream solution $(U, h)$ involved in (\ref{dispers})
and (\ref{gamma}). The last equality allows us to investigate whether equation
(\ref{dispers}) has roots. In particular, the existence of a positive root in the
absence of the above mentioned Dirichlet eigenvalues follows from positivity of the
fraction in the square brackets. On the other hand, if there is a Dirichlet
eigenvalue, then (\ref{dispers}) has at least one root greater than this eigenvalue
[see Proposition~3.3 (ii)]. Moreover, if the fraction in the square brackets is
positive, then there also exists a root between zero and this eigenvalue. Finally,
for any $r > r_c$ the equation ${\cal R}_0^{(+)} (s) = r$ has a root $s_* > s_c$. If
$\sigma$ defined by the stream solution $(U, h)$ corresponding to this $s_*$ is a
continuous function, then equation (\ref{dispers}) has no positive solutions. The
latter fact has the well-known analogue for the zero vorticity: only solitary waves
exist in the supercritical case (see, for example, \cite{AT} and \cite{KK2}).

If the function $\sigma (\tau)$ is defined for all $\tau$, then positivity of the
fraction in the square brackets is necessary and sufficient for the existence of a
root of equation (\ref{dispers}). For unidirectional shear flows this condition is
equivalent to the inequality used in \cite{KKL}, Theorem~1:
\[ \int_0^1 \frac{\D \tau}{[s^2 - 2 \Omega (\tau)]^{3/2}} > 1 \, .
\]
For flows supporting solitary waves, the left-hand side was recently considered in
\cite{Wheel}, where its reciprocal is referred to as the Froude number squared. In
the old paper \cite{Ter} by Ter-Krikorov, this number was implicitly used for the
same flows as in \cite{Wheel}. It should be added that another definition of the
Froude number suitable for both solitary and periodic waves with vorticity was
proposed and investigated by Fenton \cite{F}.

Now we turn to the second assumption.

\vspace{2mm}

\noindent (II) The dispersion equation (\ref{dispers}) has at least one positive
root, say, $\tau_0$ such that none of the values $k \tau_0$ $(k = 1,2,\dots)$ is a
root of (\ref{dispers}).

\vspace{2mm}

In section~3, we consider conditions under which this assumption is fulfilled, but
here we restrict ourselves to formulating the following crucial assertion, which is
analogous to the fact that the $\tau$-derivative of the left-hand side in
(\ref{coth}) is positive for $\tau > 0$.

\vspace{2mm}

\noindent {\bf Lemma 1.1.}  {\it On every interval of the half-axis $\tau > 0$ on
which $\sigma (\tau)$ is defined its derivative does not vanish and its sign is the
same as that of $U' (h)$, and so $\sigma (\tau)$ is monotonic there. Moreover, all
positive roots of the dispersion equation $(\ref{dispers})$ are simple.}

\vspace{2mm}

Now we define a bifurcation wavelength. Let assumptions (I) and (II) hold, and so
there exists a root $\tau_0$ of (\ref{dispers}). Then $\Lambda_0 = 2 \pi / \tau_0$
is a wavelength of linear rotational waves on the shear flow described by $(U, h)$
corresponding to $s_*$ introduced in assumption (I). If equation (\ref{dispers}) has
more than one root satisfying assumption (II), then each of these roots defines a
wavelength of linear waves with vorticity and all of them exist on the free surface
of one and the same shear flow.

Lemma 1.1 is crucial for proving the existence of local bifurcation because the
transversality condition used in the proof is equivalent to the following one:
$\sigma_\tau (\tau_0) \neq 0$.

\vspace{2mm}

Let us describe function spaces used in what follows. For $\alpha \in (0,1)$ and a
non-negative integer $k$ the space of $\Lambda_0$-periodic, even $C^{k,
\alpha}$-functions on $\RR$ is denoted by $\Pi^{k, \alpha}_{\Lambda_0}$ (cf.
\cite{KK1}, p.~478). Furthermore, let a $C^{k, \alpha}$-strip have
$\Lambda_0$-periodic and even in $X$ upper boundary. Then by $\tilde \Pi^{k,
\alpha}_{\Lambda_0}$ we denote the space of $C^{k, \alpha}$-functions on the closed
strip that are $\Lambda_0$-periodic and even in $X$. Now we are in a position to
formulate the following result.

\vspace{2mm}

\noindent {\bf Theorem 1.2.} {\it Let $\omega \in C^{2,\alpha} (\RR)$, $\alpha \in
(0,1)$, and let $r > 0$ satisfy assumption} (I). {\it If $(U, h)$ is defined by $r$
and assumption} (II) {\it holds for equation $(\ref{dispers})$ corresponding to this
stream solution, then there exist $\varepsilon > 0$ and a continuously differentiable
mapping from $\{ t \in \RR : |t| \in (0, \varepsilon) \}$ to a neighbourhood of $(0,
0)$ in $\RR \times \Pi^{2, \alpha}_{\Lambda_0}$.

The first component of this mapping is $\lambda (t) \not\equiv 0$ such that
$\lambda (t) \to 0$ as $|t| \to 0$, whereas the second component
\begin{equation}
t \, \cos \frac{2 \pi X}{\Lambda (t)} + \zeta_* \left( \frac{X}{1 + \lambda (t)} , t
\right) \ \ is \ equal \ to \ \ \xi (X, \, t) - \frac{h}{1 + \lambda (t)} \ \ for \
every \ t . \label{xi}
\end{equation}
Here $\Lambda (t) = \Lambda_0 [1 + \lambda (t)]$, $\| \zeta_* (\cdot , t)
\|_{\Pi^{2, \alpha}_{\Lambda_0}} = o (t)$ as $|t| \to 0$, and $\xi (X, \, t)$ is
the $\Lambda$-periodic upper boundary of ${\cal D}$. In the latter domain, problem
$(\ref{1})$--$(\ref{3})$ has a solution
\begin{equation}
\Psi (X, Y, t) = U \! \left( Y \frac{h}{\xi (X , t)} \right) + \Psi_* \left(
\frac{X}{1 + \lambda (t)}, \frac{Y}{1 + \lambda (t)}, t \right) \label{estim}
\end{equation}
such that $\| \Psi_* (\cdot, t) \|_{\tilde \Pi^{2, \alpha}_{\Lambda_0}} = O (t)$ as
$t \to 0$, and the pair $(\Psi (X, Y, t) , \, \xi (X , t))$ satisfies
condition~$(\ref{4})$.}

\vspace{2mm} 

This theorem means that $(\Psi, \, \xi)$ is a $\Lambda$-periodic Stokes-wave
solution of problem (\ref{1})--(\ref{4}). Moreover, $(\Psi, \, \xi)$ is a
perturbation of the stream solution $(U, \, h)$. Relation (\ref{xi}) immediately
yields this for $\xi$, whereas for $\Psi$ this fact will be shown in the theorem's
proof, for which purpose the form of function $\Psi_*$ will be specified (see
section~4.2). Moreover, $(\Psi, \, \xi)$ belongs to a family of solutions describing
Stokes waves with wavelengths close to $\Lambda_0$ provided the latter is defined by
a root of the dispersion equation (\ref{dispers}). If $r_0$ (see formula (\ref{ss4})
in Appendix B for its definition) is finite and $r > r_0$, then $(U, \, h)$
describes a shear flow with counter-currents (see \cite{KK} and section 6, where
examples are considered).

Note that a result similar to Theorem 1.2 is true when $s_* = s_0 = 0$, $r_0 <
\infty$ and $U' (h) \neq 0$. This case is illustrated by Wahl\'en in \cite{Wah}; see
the pattern of streamlines plotted in Figure~2 of his paper for the case when
$\omega$ is a negative constant.

It is worth emphasizing that the mechanism of wave bifurcation is substantially more
complicated for waves with vorticity comparing with that for irrotational waves.
First, for every $r > r_c$ there are exactly two uniform flows (sub- and
supercritical) in the irrotational case, whereas there can be as many rotational
flows of constant depth as one pleases which, for example, is the case for linear
positive vorticity (see section~6.2 below). Second, for a given depth of the
subcritical irrotational uniform flow the dispersion equation (\ref{coth}) defines
only a single value of the bifurcation wavelength. On the other hand, the number of
roots of the rotational dispersion equation (\ref{dispers}) depends on the value of
depth (the second component of the stream solution), which, in its turn, depends on
$r > r_c$. Moreover, the number of such depths can be as large as one pleases, which
again is the case for linear positive vorticity.

\subsection{Bifurcation with fixed wavelength: formulation of main results}

Another option for Stokes waves bifurcating from a horizontal shear flow is to keep
the wavelength fixed. Since all stream solutions are parameterised by $s \geq s_0$
(see formulae (\ref{s_0}) for the definition of $s_0$), it is reasonable to base a
bifurcation parameter on $s$. According to results obtained in \cite{KK4}, no steady
waves of small amplitude are supported by a shear flow corresponding to $s = s_0$
and having a still free surface. Therefore, values of $s$ must be strictly greater
than $s_0$ unless the latter is equal to zero and $U' (h) \neq 0$ for the
corresponding stream solution.

Thus, we take some $s_* > s_0$ as the starting point; $s_*$ exists because $r > r_c$
by assumption (I). Let the corresponding stream solution be $(U (\cdot; s_*), \, h
(s_*))$, where $h$ is given by one of formulae (\ref{h2k})--(\ref{hj-}). In what
follows, this stream solution is denoted $(U, h)$ in order to distinguish it from
$(U (\cdot; s), \, h (s))$ with $s \neq s_*$. To be specific, we suppose that $U'
(h) > 0$; the case when the opposite inequality holds needs only elementary
amendments (cf. section 2.2 below). Using assumption (II), we choose some positive
root $\tau_0$ of the dispersion equation (\ref{dispers}), but now we keep the
wavelength $\Lambda_0 = 2 \pi / \tau_0$ fixed, whereas $s$ varies near $s_*$. Thus,
the problem's small parameter $\mu = s - s_*$ appears in the problem through the
corresponding stream solution and the right-hand side in the Bernoulli equation
(\ref{4}) which is equal to
\[ r (s_* + \mu) = r (s) = \frac{1}{3} \left[ s^2 - 2 \, \Omega (1) + 2 \, h (s)
\right] , 
\]
[cf. the second formula (\ref{ss2})].

Similar to (\ref{gamma}) we consider $\gamma_* (Y, \tau; s)$ that solves the
problem
\begin{equation}
- \gamma''_* + [ \tau^2 - \omega' (U (Y; s)) ] \, \gamma_* = 0 \ \mbox{on} \ (0, h
(s)), \quad \gamma_* (0, \tau; s) = 0 , \ \ \gamma_* (h (s), \tau; s) = 1 .
\label{gamma_*}
\end{equation}
In terms of $\gamma_* (Y, \tau; s)$ and $\kappa (s) = U' (h (s); s)$ [in other
words, $\kappa (s) = [3 \, r (s) - 2 \, h (s)]^{1/2}$] we introduce the following
dispersion equation:
\begin{equation}
\sigma_* (\tau; s) = 0 , \ \ \mbox{where} \ \sigma_* (\tau; s) = \kappa (s) \,
\gamma'_* (h (s), \tau; s) - [\kappa (s)]^{-1} + \omega (1) . \label{dispers_*}
\end{equation}
Note that $\sigma_* (\tau; s_*) = \sigma (\tau)$, in which case equation
(\ref{dispers_*}) coincides with (\ref{dispers}). Along with (I) and (II) we will
use the following assumption.

\vspace{2mm}

\noindent (III) If $\tau_0$ is a root of the dispersion equation (\ref{dispers_*})
with $s = s_*$ [this root exists by assumption (II)], then $\dot{\sigma}_* (\tau_0;
s_*) \neq 0$.

\vspace{2mm}

\noindent Here and below the top dot denotes differentiation with respect to $s$.
Now we are in a position to formulate the following alternative to Theorem 1.2.

\vspace{2mm}

\noindent {\bf Theorem 1.3.} {\it Let $\omega \in C^{2,\alpha} (\RR)$, $\alpha \in
(0,1)$. If assumptions} (I)--(III) {\it hold for $r = r (s_*)$ and the stream
solution $(U, h)$ corresponding to $r (s_*)$, then there exist $\varepsilon > 0$ and
a continuously differentiable mapping from $\{ t \in \RR : |t| \in (0, \varepsilon)
\}$ to a neighbourhood of $(0, 0)$ in $\RR \times \Pi^{2, \alpha}_{\Lambda_0}$.

The first component of this mapping is $\mu (t) \not\equiv 0$ such that $\mu (t)
\to 0$ as $|t| \to 0$, whereas the second component
\[ t \, \cos \frac{2 \pi X}{\Lambda_0} + \zeta_\circ (X, t) \ \ is \ equal \ to \ \ 
\xi (X, \, t) - h (s_* + \mu (t)) \ \ for \ every \ t .
\]
Here $\| \zeta_\circ (\cdot , t) \|_{\Pi^{2, \alpha}_{\Lambda_0}} = o (t)$ as $|t|
\to 0$ and $\xi (X, \, t)$ is the $\Lambda_0$-periodic upper boundary of ${\cal D}$.
In the latter domain, problem $(\ref{1})$--$(\ref{3})$ with $r = r (s_* + \mu (t))$
has a solution $\Psi (\cdot , t) \in \tilde \Pi^{2, \alpha}_{\Lambda_0}$ such that
the pair $(\Psi (X, Y, t) , \, \xi (X , t))$ satisfies condition $(\ref{4})$.
Moreover,}
\[ \Psi (X, Y, t) = U \! \left( Y \frac{h (s_* + \mu (t))}{\xi (X , t)} \right) 
+ \Psi_\circ (X, Y, t) , \ \ where \ \| \Psi_\circ (\cdot, t) \|_{\tilde \Pi^{2,
\alpha}_{\Lambda_0}} = O (t) \ as \ t \to 0. \]

This theorem describes a bifurcation mechanism alternative to that presented in
section~1.4. In this connection, it worth to mention the result obtained in
section~6.3. It says that there exists a family of shear flows depending on $s$ in
such a way that for some $s_\circ > s_0$ the corresponding equation
(\ref{dispers_*}) has a root $\tau (s_\circ)$ for which (III) is violated, that is,
$\sigma_* (\tau (s_\circ); s_\circ) = 0$  and $\dot{\sigma}_* (\tau (s_\circ);
s_\circ) = 0$ hold simultaneously. This provides a possibility that the mechanism
described in Theorem~1.3 fails to yield a brunch of Stokes waves that has the
constant wavelength defined by $s_\circ$ and bifurcates from the corresponding shear
flow. Since Theorem 1.2 guarantees that a brunch of Stokes waves with varying
wavelength bifurcates from this shear flow, the latter bifurcation mechanism might
be more flexible than that described in Theorem~1.3.

\section{Operator form of problem (\ref{1})--(\ref{4}) \\ with fixed Bernoulli's
constant}

We prove Theorem 1.2 using the Crandall--Rabinowitz theorem (see Theorem~4.1 in
section~4), to apply which to problem (\ref{1})--(\ref{4}) the latter must be
transformed into a nonlinear operator equation.

\subsection{Reformulation of the problem}

First, we choose a horizontal shear flow from which Stokes waves bifurcate. In view
of assumption (I) we fix $r > r_c$ (the critical value $r_c$ defined by formula
(\ref{ss3}) depends on the vorticity distribution), and so equation (\ref{ss2}) has
a solution $s_*$ at least for one function ${\cal R}_j^{(\pm)}$ $(j=0,1,\dots)$.
The stream solution corresponding to $s_*$ is as follows:
\begin{equation}
\big( U_j^{(\pm)} (Y;s_*), \, h_j^{(\pm)} (s_*) \big) . \label{sol}
\end{equation} 
Here the second component is given by one of the formulae (\ref{h2k})--(\ref{hj-})
and expressions for the first component can be found in \cite{KK}, section~3,
whereas its properties are described above in section~1.3. For the sake of brevity,
the chosen solution will be denoted $(U, h)$ in what follows, whereas ${\cal R}$
will stand for the function ${\cal R}_j^{(\pm)}$ defining $(U, h)$. Note that $U'
(h, s_*) \neq 0$, which is also a consequence of assumption (I). If for some $r >
r_c$ equation (\ref{ss2}) has roots for several functions ${\cal R}_j^{(\pm)}$, then
each of these roots can be used for obtaining the corresponding dispersion equation
with the help of solution (\ref{sol}). Moreover, each of these dispersion equations
can have one or more roots, and so there can exist more than one bifurcation
wavelength.

\vspace{2mm}

Now we use an appropriate scaling in order to reformulate the problem and to
introduce a small parameter. Let $\Lambda$ denote the wavelength of Stokes waves
perturbing the free surface $Y = h$ of the chosen shear flow. Assuming that
$\Lambda$ is close to a certain bifurcation wavelength $\Lambda_0 > 0$, we put
$\Lambda = \Lambda_0 (1+\lambda)$, and so $\lambda$ is a small parameter to be found
along with the stream function and the wave profile. It will be shown below (see
Proposition~2.1) that $\Lambda_0 = 2 \pi / \tau_0$ gives a bifurcation wavelength
provided $\tau_0$ is a root of the dispersion equation (\ref{dispers}).

For transforming problem (\ref{1})--(\ref{4}) into a new one depending on
$\lambda$, we introduce the following variables:
\begin{equation}
x = \frac{X \Lambda_0}{\Lambda}, \quad y = \frac{Y \Lambda_0}{\Lambda}, \quad \eta
(x) = \frac{\xi (X) \Lambda_0}{\Lambda}, \quad \psi (x,y) = \Psi (X,Y) \, .
\label{nv}
\end{equation}
Then instead of ${\cal D}$ we get the curved strip $D = \{ -\infty < x < +\infty ,\
0 < y < \eta (x) \}$ and the problem takes the form:
\begin{eqnarray}
&& \psi_{xx} + \psi_{yy} + (1+\lambda)^2 \omega (\psi) = 0,\quad (x,y) \in D ;
\label{1'} \\ && \psi (x, 0) = 0,\quad x \in \RR; \label{2'} \\ && \psi (x, \eta (x))
= 1, \quad x \in \RR; \label{3'} \\ && |\nabla_{x,y} \psi (x, y)|^2 + 2
(1+\lambda)^3 y = 3 r (1+\lambda)^2 , \quad y = \eta (x), \ x \in \RR . \label{4'}
\end{eqnarray}
Now we seek $(\psi, \eta)$ and $\lambda$, so that $\eta$ is a non-constant,
$\Lambda_0$-periodic, even function, whereas $\psi$ is $\Lambda_0$-periodic and even
in $x$.

In view of the boundary condition (\ref{3'}), equation (\ref{4'}) takes the form:
\begin{equation}
\frac{\partial \psi}{\partial n} \, (x, \eta (x)) - (1+\lambda) \left[ 3r - 2
(1+\lambda) \, \eta (x) \right]^{1/2} = 0 , \quad x \in \RR , \label{beq}
\end{equation}
provided $U' (h) > 0$; here $n$ is the unit normal directed outwards of $D$. If $U'
(h) < 0$, then we have
\begin{equation}
\frac{\partial \psi}{\partial n} \, (x, \eta (x)) + (1+\lambda) \left[ 3r - 2
(1+\lambda) \, \eta (x) \right]^{1/2} = 0 , \quad x \in \RR  \label{beq-}
\end{equation}
instead of (\ref{beq}). In what follows, we restrict our considerations to
(\ref{beq}), just formulating the results that concern (\ref{beq-}).

\subsection{Reduction to an operator equation}

The first step of our reduction of problem (\ref{1'})--(\ref{4'}) to an operator
equation is to transform the curved strip $D$ into $S = \RR \times (0, h)$. For
this purpose we change the vertical coordinate $y$ to the following one:
\begin{equation}
z = y \frac{h}{\eta (x)}  \quad \left[ \mbox{or} \ z = Y \frac{h}{\xi (X)} \
\mbox{according to (\ref{nv})} \right] . 
\label{z}
\end{equation}
In what follows, the first component of the chosen solution to problem (\ref{ss1})
will be considered as a function of $z$ and denoted by $u (z)$. (In section~1.3, we
described how to find this stream solution $(U, h)$ given by formula (\ref{sol}).)
Second, we define new unknown functions $\zeta (x)$ on $\RR$ and $\phi (x, z)$ on
$\bar S$ as follows:
\begin{equation}
\zeta (x) = \eta (x) - \frac{h}{1 + \lambda} \quad \mbox{and} \quad  \phi (x, z) =
\psi \left( x, \frac{z}{h} \, \eta (x) \right) - u (z) . \label{zeta/phi}
\end{equation}
Thus $\zeta$ and $\phi$ are $\Lambda_0$-periodic and even in $x$ functions, but
they are small when a perturbed flow is close to that defined by the pair $(u, h)$,
which is nothing else than the stream solution $(U, h)$ written in the variables
$(x, z)$.

\vspace{2mm}

Finally, we have to describe how $\phi$ is related to $\lambda$ and $\zeta$, for
which purpose we use the weak setting of problem (\ref{1'})--(\ref{3'}); the
corresponding integral identity is as follows:
\begin{equation} 
\int_{\cal D} [ \psi_x v_x + \psi_y v_y - (1+\lambda)^2 \omega (\psi) \, v ] \, \D x
\, \D y = 0 . \label{iipsi}
\end{equation}
It is obtained from equation (\ref{1'}) by integration by parts and must be valid
for all $v \in W^{1,2}_{\rm loc} ({\cal D})$ vanishing on $\partial {\cal D}$;
$W^{1,2}_{\rm loc} ({\cal D})$ denotes the space each element of which belongs to
$W^{1,2} (K)$ for some bounded open subset $K \subset {\cal D}$.

In identity (\ref{iipsi}), we change the variables $(x,y)$ and the function $\psi$
to $(x,z)$ and $\phi$, respectively. Moreover, we take into account the
$\Lambda_0$-periodicity and evenness of $\eta$, and the fact that $\phi$ is also
periodic and even function of $x$, which allows us to integrate over $(-\Lambda_0/2,
\Lambda_0/2)$ instead of $\RR$. (It is convenient to take $(-\Lambda_0/2 ,
\Lambda_0/2)$ as the periodicity interval.) Therefore, taking into account the first
relation (\ref{ss1}) and the fact that $\eta_x = \zeta_x$, we arrive at the following
identity:
\begin{eqnarray}
\int_{-\Lambda_0/2}^{\Lambda_0/2} \int_0^h \Bigg\{ \left[ \phi_x - \frac{z \zeta_x}
{\eta} (u_z + \phi_z) \right] \left[ v_x - \frac{z \zeta_x}{\eta} v_z \right] +
\left( \frac{h}{\eta} \right)^2 \phi_z v_z \nonumber \\ + \left[ \left(
\frac{h}{\eta} \right)^2 \omega (u) - (1+\lambda)^2 \omega (u + \phi) \right] v
\Bigg\} \frac{\eta \, \D x \, \D z}{h} = 0 \, . \label{iiphi'}
\end{eqnarray}
It is valid for all $v \in W^{1,2} (S)$ vanishing on $\partial S$ and
$\Lambda_0$-periodic in $x$.

Now we are in a position to derive an operator equation for $\Phi = (\phi, \zeta)$.
First, we apply the divergence theorem to (\ref{iiphi'}) and, in view of
arbitrariness of the test function $v$, obtain that
\begin{equation}
F_1 (\Phi; \lambda) = 0 , \label{nov1}
\end{equation}
where
\begin{eqnarray}
&& F_1 (\Phi; \lambda)= - \bigg\{ \frac{\eta}{h} \left[ \phi_x - \frac{z
\zeta_x}{\eta} (u_z + \phi_z) \right] \bigg\}_x + \bigg\{ \frac{z \zeta_x}{h} \left[
\phi_x - \frac{z \zeta_x}{\eta} (u_z + \phi_z) \right] \bigg\}_z \nonumber \\ && \ \
\ \ \ \ \ \ \ \ \ \ \ \ \ \ \ \ - \frac{h}{\eta} \, \phi_{zz} + \frac{\eta}{h}
\left[ \left( \frac{h}{\eta} \right)^2 \omega (u) - (1+\lambda)^2 \omega (u + \phi)
\right] . \label{nov2}
\end{eqnarray}
Here $(x, y) \in (-\Lambda_0/2, \Lambda_0/2) \times (0, h)$; $\lambda$ is to be
found along with $\Phi$, whereas $\eta = \zeta + h / (1+\lambda)$ [see the first
formula (\ref{zeta/phi})]. Second, we apply (\ref{zeta/phi}) and get
\[ \frac{\partial \psi}{\partial n} \, (x, \eta (x)) = \left[ u_z (h) + \phi_z (x, h)
\right] \frac{h \sqrt{1+\eta_x^2 (x)}}{\eta (x)} - \frac{\phi_x (x, h) \, \eta_x
(x)}{\sqrt{1+\eta_x^2 (x)}} \, . 
\]
Furthermore, $\phi_x (x, h)$ vanishes identically, and so (\ref{beq}) takes the form:
\[ \left[ u_z (h) + \phi_z (x, h) \right] \frac{h}{\eta (x)} - (1+\lambda) \left[
\frac{3r - 2 (1+\lambda) \, \eta (x)}{1+\eta_x^2 (x)} \right]^{1/2} = 0 , \quad x
\in \RR .
\]
As above $\eta = \zeta + h / (1+\lambda)$. Finally, we use that $u_z (h) = \kappa$
(it is equal to $(3r - 2h)^{1/2}$ provided $u_z (h) > 0$), and obtain after simple
algebra the following form of Bernoulli's equation:
\begin{eqnarray}
&& \ \ \ \ \ \ \ \ \ \ \ \ \ \ \ \ \ \ \ \ \ \ \ \ \ \ \ \ \ F_2 (\Phi; \lambda) = 0
, \quad \mbox{where} \label{nov3} \\ && F_2 (\Phi; \lambda) = \phi_z (x, h) -
\left[ 1 + \frac{1 + \lambda}{h} \, \zeta (x) \right] \left[ \frac{\kappa^2 - 2
(1+\lambda) \, \zeta (x)}{1+\zeta_x^2 (x)} \right]^{1/2} + \kappa \, . \label{beq'}
\end{eqnarray}
Here $x \in (-\Lambda_0/2, \Lambda_0/2)$ and the values of $h$ and $\kappa$ are
given (both of them are functions of $r$), whereas $\Phi$ and $\lambda$ are unknown.
We recall that (\ref{beq'}) is obtained under the assumption that $U' (h) > 0$.
Supposing that $U' (h) < 0$, we have $\kappa = u_z (h) = -(3r - 2h)^{1/2}$, and so
\begin{equation}
F_2 (\Phi; \lambda) = \phi_z (x, h) + \left[ 1 + \frac{1 + \lambda}{h} \, \zeta (x)
\right] \left[ \frac{\kappa^2 - 2 (1+\lambda) \, \zeta (x)}{1+\zeta_x^2 (x)}
\right]^{1/2} + \kappa , \label{beq'-}
\end{equation}
instead of (\ref{beq'}); here again $x \in (-\Lambda_0/2, \Lambda_0/2)$.

Combining (\ref{nov1}) and (\ref{nov3}), we write problem (\ref{1})--(\ref{4}) as
the following operator equation
\begin{equation}
F (\Phi; \lambda) = 0 , \quad \mbox{where} \ F = (F_1, F_2)  \label{nov4}
\end{equation}
is the nonlinear operator whose components $F_1$ and $F_2$ are given by formulae
(\ref{nov2}) and (\ref{beq'}), respectively, provided $U' (h) > 0$ [if $U' (h) < 0$,
then (\ref{beq'-}) must be used instead of (\ref{beq'})]. Here $F$ is considered as
acting in the following function space. Let $\Pi^{k, \alpha}_{\Lambda_0} (\bar S)$
[$k$ is a non-negative integer and $\alpha \in (0,1)$] be the space consisting of
$C^{k, \alpha}$-functions on $\bar S$ that are $\Lambda_0$-periodic and even in $x$.
Putting ${\cal X}^{k, \alpha} = \Pi^{k, \alpha}_{\Lambda_0} (\bar S) \times \Pi^{k,
\alpha}_{\Lambda_0}$ and denoting by ${\cal X}_0^{2, \alpha}$ its subspace
consisting of elements whose first components vanish for $z=0$ and $z=h$, we see
that $F$ maps ${\cal X}_0^{2, \alpha} \times (-\delta, \delta) \mapsto {\cal X}^{0,
\alpha}$ continuously; here $\delta$ is a sufficiently small positive number.

\subsection{The Fr\'echet derivative of $F$ and the dispersion equation
(\ref{dispers})}

It is straightforward to find the Fr\'echet derivative $F_\Phi (0;0)$; being applied
to $(\hat \phi, \hat \zeta) \in {\cal X}_0^{2, \alpha}$, it has the following
components:
\begin{eqnarray}
&& - \hat \phi_{xx} - \hat \phi_{zz} - \omega' (u) \, \hat \phi - h^{-1} [ - z \,
u_z \, \hat \zeta_{xx} + 2 \, \omega (u) \, \hat \zeta ] \, ,  \label{nov5} \\ && \
\ \ \ \ \ \ \ \ \ \ \ \ \ \ \ \ \ \ \left[ \hat \phi_z \right]_{z=h} - \left(
\frac{\kappa}{h} - \frac{1}{\kappa} \right) \hat \zeta . \label{nov6}
\end{eqnarray}
The kernel of this operator is described in the next assertion.

\vspace{2mm}

\noindent {\bf Proposition 2.1.} {\it A pair $(\hat \phi, \hat \zeta) \in {\cal X}_0^{2, \alpha}$ satisfies the
equation $[F_\Phi (0;0)] \, (\hat \phi, \hat \zeta) = 0$ if and only if either
\begin{equation}
\hat \phi (x, z) = W (z) \cos \tau x \quad and \quad \hat \zeta (x) = \varsigma
\cos \tau x , \label{nov7}
\end{equation}
where $\tau = 2 \pi / \Lambda_0$ is a root of $(\ref{dispers})$, $\varsigma$ is a
non-zero constant and $W$ is a non-zero function on the interval $(0, h)$ such that
\begin{equation}
- W_{zz} + [ \tau^2 - \omega' (u) ] \, W = [ z \, u_z \, \tau^2 + 2 \, \omega (u) ]
(\varsigma / h) \ \ \mbox{on} \ (0, h) , \quad W (0) = W (h) = 0 ,
\label{nov9}
\end{equation}
and
\begin{equation}
W_z (h) = \left( \frac{\kappa}{h} - \frac{1}{\kappa} \right) \varsigma ,
\label{nov10}
\end{equation}
or $\hat \phi$ and $\hat \zeta$ are linear combinations of functions of the form
$(\ref{nov7})$ corresponding to different roots of $(\ref{dispers})$, namely, $\tau
= 2 \pi k / \Lambda_0$ with integer values $k > 0$.}

\vspace{2mm}

Note that formulating this proposition we do not suppose assumption (II) to be
fulfilled. Prior to proving the proposition we prove the following assertion.

\vspace{2mm}

\noindent {\bf Lemma 2.2.} {\it Problem $(\ref{nov9})$ with $\varsigma \neq 0$ is
solvable if and only if the boundary value problem
\begin{equation}
- w_{zz} + [ \tau^2 - \omega' (u) ] \, w = 0 \ \ \mbox{on} \ (0, h) , \quad w (0) = w
(h) = 0 \label{nov8}
\end{equation}
has only a trivial solution.}

\vspace{2mm}

\noindent {\it Proof.} Let us assume that problem (\ref{nov8}) has a non-trivial
solution. Then it must be orthogonal to the right-hand side of equation (\ref{nov9})
because $W$ is a non-zero function, that is, we have
\begin{equation}
\varsigma \int_0^h [ z \, u_z \, \tau^2 + 2 \, \omega (u) ] \, w \, \D z = 0 .
\label{nov11}
\end{equation}
Since $\omega (u) = - u_{zz}$, $\tau^2 \, w = w_{zz} + \omega' (u) \, w$ on the
interval $(0, h)$ and $\omega' (u) u_z = - u_{zzz}$, the integral can be written as
follows:
\begin{eqnarray*}
&& \int_0^h \{ z \, u_z \, [ w_{zz} + \omega' (u) \, w ] - 2 \, u_{zz} \, w \} \, \D
z \\ && = \int_0^h ( z \, u_z \, w_{zz} - z \, w \, u_{zzz} - 2 \, u_{zz} \, w ) \,
\D z \\ && = \varsigma \int_0^h [ z ( u_z \, w_{z} )_z - u_{zz} \, w ] \, \D z = h
\, u_z (h) \, w_z (h) ,
\end{eqnarray*}
where the last equality follows by integration by parts in both terms. In view of
(\ref{nov11}) we get that $\varsigma \, u_z (h) \, w_z (h) = 0$. Since $\varsigma
\neq 0$, $u_z (h) \neq 0$ by assumption (I) and $w_z (h) \neq 0$ because otherwise
$w$ vanishes identically as a solution of the homogeneous Cauchy problem, we have a
contradiction which proves the lemma's assertion.

\vspace{2mm}

\noindent {\it Proof of Proposition 2.1.} Equating expressions (\ref{nov5}) and
(\ref{nov6}) to zero, we apply separation of variables to the obtained problem, thus
finding that all its solutions are linear combinations of pairs (\ref{nov7}) such
that relations (\ref{nov9}) and (\ref{nov10}) are fulfilled, whereas $\tau = 2 \pi k
/ \Lambda_0$ with a positive integer $k$.

Let us multiply equation (\ref{nov9}) by
$\gamma (z, \tau)$ and integrate over $(0, h)$. After integration by parts twice in
the left-hand side of the resulting equality we obtain that the integral term
cancels in view of the equation for $\gamma$, whereas the boundary conditions for
$W$ and $\gamma$ yield the relation
\begin{equation}
W_z (h) = - \sigma_0 (\tau) \, \varsigma , \quad \mbox{where} \ \ \sigma_0
(\tau) = h^{-1} \int_0^h \gamma (z, \tau) \, [ \tau^2 z u_z + 2 \, \omega (u (z)) ]
\, \D z . \label{sigma}
\end{equation}
Comparing this and (\ref{nov10}), we get that $\tau$ must satisfy the following form
of the dispersion equation:
\begin{equation}
\sigma_0 (\tau) + \frac{\kappa}{h} - \frac{1}{\kappa} = 0 . \label{nov18}
\end{equation}
In order to calculate the integral in (\ref{sigma}) we proceed in the same way as
for the integral (\ref{nov11}), thus getting that
\[ \sigma_0 (\tau) = \frac{1}{h} \int_0^h \left( z \gamma_{zz} u_{z} - 
z \gamma \, u_{zzz} - 2 \, \gamma \, u_{zz} \right) \D z .
\]
Integrating by parts in the middle term, we obtain
\[ \sigma_0 (\tau) = - u_{zz} (h) + \frac{1}{h} \int_0^h \left[ z \left( \gamma_z u_z
\right)_z - \gamma \, u_{zz} \right] \D z = \omega (1) + \kappa \, \gamma_z (h) -
\frac{\kappa}{h} ,
\]
where it is taken into account that $- u_{zz} (h) = \omega (1)$ and $u_z (h) =
\kappa$. Hence (\ref{nov18}) coincides with (\ref{dispers}).

\vspace{2mm}

\noindent {\bf Remark 2.3.} According to Proposition 2.1, the value $\Lambda_0$
introduced at the beginning of section~2.1 serves as a bifurcation wavelength when
$\Lambda_0 = 2 \pi / \tau_0$ and $\tau_0$ is a root of the dispersion equation
(\ref{dispers}).

\subsection{Proof of Lemma 1.1} 

Differentiating relations (\ref{gamma}) with respect to $\tau$, we obtain that
$\gamma_\tau (Y, \tau)$ solves the following problem:
\[ - \gamma_\tau'' + [ \tau^2 - \omega' (U) ] \, \gamma_\tau = - 2 \, \tau \gamma \
\mbox{on} \ (0, h), \quad \gamma_\tau (0, \tau) = 0 , \ \ \gamma_\tau (h, \tau) = 0
,
\]
which is similar to problem (\ref{nov9}), and so we apply the considerations used
for proving Lemma 2.2. Namely, we multiply the last equation by $\gamma$, integrate
by parts twice in the left-hand side and use the equation and boundary conditions
for $\gamma$ and the boundary conditions $\gamma_\tau$. This yields that
$\gamma'_\tau (h, \tau) = 2 \, \tau \int_0^h \gamma^2 (z, \tau) \, \D z$. On the
other hand, it follows from the definition of $\sigma$ that $\sigma_\tau (\tau) =
\kappa \, \gamma'_\tau (h, \tau)$. Combining the last two equalities we arrive at
the proposition's assertion.

\section{On roots of the dispersion equation (\ref{dispers})}

In this section, we consider the dispersion equation (\ref{dispers}) corresponding
to some stream solution. First, we prove sufficient conditions (some of them are
also necessary) which guarantee that (\ref{dispers}) has at least one positive root,
and so assumption (II) is fulfilled. Second, we show that (\ref{dispers}) has no
roots in the case when this equation corresponds to a stream solution defined by a
{\it supercritical} value of $s$, that is, $s$ is greater than $s_c$ and satisfies
the equation ${\cal R}_0^{(+)} (s) = r$, where $r$ is any number greater than $r_c$.
The latter fact is analogous to that well known for zero vorticity, namely, only
solitary waves exist in the supercritical case (see, for example, \cite{AT} and
\cite{KK2}).

\subsection{Conditions of solvability of equation (\ref{dispers})}

Let assumption (I) hold, and so there exists $s_*$ that solves equation (\ref{ss2})
for some ${\cal R}$, and let $(U, h)$ be the stream solution corresponding to $s_*$.
(Here we use the simplified notation introduced above, namely, $(U, h)$ stands for
the stream solution, whereas ${\cal R}$ denotes the left-hand side of equation
(\ref{ss2}) whose root $s_*$ defines this solution.) We begin with two auxiliary
assertions concerning the function $\sigma$ defined by $(U, h)$. The first of them
describes the behaviour of $\sigma (\tau)$ at infinity, and the second assertion
gives the asymptotics of $\gamma (Y, \tau)$ near an eigenvalue (under the assumption
that it exists) of the operator $\D^2 / \D Y^2 + \omega' (U)$ with the Dirichlet
boundary conditions. (We recall that only a finite number of such eigenvalues can
exist.) Then we prove the proposition about solvability of equation
(\ref{dispers}).

\vspace{2mm}

\noindent {\bf Lemma 3.1.} {\it The following asymptotic formula holds}
\begin{equation}
\sigma (\tau) = |\tau| \, U' (h) + O (1) \quad as \ \tau \to +\infty .
\label{inf<}
\end{equation}

\noindent {\it Proof.} In order to prove this formula we consider the solution
$\gamma (Y, \tau)$ of the boundary value problem (\ref{gamma}) and investigate its
asymptotic behaviour as $\tau \to +\infty$. Note that $\gamma (Y, \tau)$ is a
smooth function of both variables for large values of $\tau$, because the operator
$\D^2 / \D Y^2 + \omega' (U)$ has only a finite number of the Dirichlet eigenvalues.
Let us write
\[ \gamma (Y, \tau) = \frac{\sinh Y \tau}{\sinh h \tau} + \tilde \gamma (Y, \tau) ,
\]
and find the asymptotics of $\tilde \gamma (Y, \tau)$ for large $|\tau|$. Since this
function satisfies the following problem:
\[ - \tilde \gamma'' + [ \tau^2 - \omega' (U) ] \, \tilde \gamma = \omega' (U) 
\frac{\sinh Y \tau}{\sinh h \tau} \ \ \mbox{for} \ Y \in (0, h) , \quad \tilde
\gamma (0) = 0 , \ \ \tilde \gamma (h) = 0 ,
\]
we have
\[ \int_0^h \tilde (\gamma')^2 \, \D Y + \tau^2 \int_0^h \tilde \gamma^2 \, \D Y -
\int_0^h \omega' (U (Y)) \, \tilde \gamma^2 \, \D Y = \int_0^h \omega' (U (Y))
\frac{\sinh Y \tau}{\sinh h \tau} \, \tilde \gamma \, \D Y \, .
\]
This is obtained by multiplying the equation by $\tilde \gamma$, integrating over
$(0, h)$ and using the boundary conditions after integration by parts.

Let $C_\omega$ bounds $\omega'$ from above, then the last equality yields that
\begin{equation} 
\int_0^h \left[ (\tilde \gamma')^2 + \frac{\tau^2}{2} \tilde \gamma^2 \right] \D Y
\leq \frac{\tau^2}{4} \int_0^h \tilde \gamma^2 \, \D Y + \frac{4 \,
C_\omega^2}{\tau^2 \, \sinh^2 h \tau} \int_0^h \sinh^2 Y \tau \, \D Y ,
\label{prev}
\end{equation}
provided $\tau$ is sufficiently large, and so the following estimate holds
\begin{equation} 
\tilde \gamma (Y, \tau) = O (\tau^{-2}) \quad \mbox{as} \ |\tau| \to \infty .
\label{gam/inf}
\end{equation}
Indeed, inequality (\ref{prev}) gives
\[ \int_0^h \left[ (\tilde \gamma')^2 + \frac{\tau^2}{4} \tilde \gamma^2 \right] 
\D Y \leq \frac{2 \, C_\omega^2}{\tau^3} \, ,
\]  
and so
\[ \max_{Y \in [0,h]} \tilde \gamma^2 (Y, \tau) \leq 2 \left( \int_0^h (\tilde 
\gamma')^2 \, \D Y \right)^{1/2} \left( \int_0^h \tilde \gamma^2 \, \D Y
\right)^{1/2} \leq \frac{8 \, C_\omega^2}{\tau^4}
\]
for large values of $|\tau|$. The last inequality immediately yields
(\ref{gam/inf}).

Furthermore, we have
\[ \frac{\sinh Y \tau}{\sinh h \tau} = \E^{|\tau| (Y-h)} + O \left( 
\E^{-|\tau| h} \right) \quad \mbox{as} \ |\tau| \to \infty ,
\]
which combined with (\ref{gam/inf}) gives that
\[ \gamma (Y, \tau) = \E^{|\tau| (Y-h)} + O (\tau^{-2}) \quad \mbox{as} \ |\tau| 
\to \infty .
\] 
Substituting this into the formula for $\sigma$ obtained in section~2.4 [its
principal integral term is given by the second relation (\ref{sigma})], we get
\[ \sigma (\tau) = \frac{\tau^2}{h} \int_0^h \E^{|\tau| (Y-h)} Y \, U' (Y) \, 
\D Y + O (1) \quad \mbox{as} \ |\tau| \to \infty .
\] 
The asymptotic formula (\ref{inf<}) is a direct consequence of this representation.

\vspace{2mm}

Now we turn to the assertion about the behaviour of $\sigma (\tau)$ when $\tau^2$ is
close to a Dirichlet eigenvalue of the operator $\D^2 / \D Y^2 + \omega' (U)$. Prior
to formulating the result, we notice that every such eigenvalue (provided it
exists) is simple.

\vspace{2mm}

\noindent {\bf Lemma 3.2.} {\it Let the operator $\D^2 / \D Y^2 + \omega' (U)$
considered on $(0, h)$ have $\tau^2_*$ as a non-zero Dirichlet eigenvalue. Then the
following asymptotic formula holds
\begin{equation}
\sigma (\tau) = \frac{- \kappa [\gamma'_* (h)]^2}{2 \tau_* (\tau - \tau_*)} + O (1)
\quad as \ \tau \to \tau_* . \label{3.2.2}
\end{equation}
Here $\gamma_* (Y)$ is the corresponding eigenfunction normalized in $L^2 (0, h)$,
and $\gamma'_* (h) \neq 0$.}

\vspace{2mm}

\noindent {\it Proof.} The constant $\gamma'_* (h)$ is not equal to zero because
otherwise the function $\gamma_*$ has the zero Cauchy data at $Y = h$, and so
vanishes identically on $(0, h)$ which is impossible.

It is clear that $\tau^2$ is not an eigenvalue of $\D^2 / \D Y^2 + \omega' (U)$
provided $\tau$ is sufficiently close to $\tau_*$. Let $\gamma (Y, \tau)$ be a
solution of problem (\ref{gamma}) for such a value of $\tau$. If we show that
\begin{equation}
\gamma (Y, \tau) = \frac{- \gamma'_* (h) \, \gamma_* (Y)}{2 \tau_* (\tau - \tau_*)}
+ v (Y, \tau) \quad \mbox{as} \ \tau \to \tau_* , \label{3.2.0}
\end{equation}
then (\ref{3.2.2}) follows from this formula in view of the definition of $\sigma$
[see (\ref{dispers}) and (\ref{gamma})]. Notice that the first term in the
right-hand side is invariant under changing of the sign of $\gamma_*$ and the
remainder $v (Y, \tau)$ is a smooth function of both variables.

In order to prove the asymptotic formula (\ref{3.2.0}) we use the following
representation
\[ \gamma (Y, \tau) = \frac{C \gamma_* (Y)}{\tau^2 - \tau_*^2} + v (Y, \tau)
\]
for $\tau$ close to $\tau_*$; here $C$ is a non-zero constant. In order to find $C$
we substitute the right-hand side into (\ref{gamma}) and get that $v$ must satisfy
the following problem:
\begin{equation}
- v'' + [ \tau^2 - \omega' (U) ] \, v = - C \gamma_* , \quad v (0) = 0 , \ \ v (h)
= 1 . \label{3.2.1}
\end{equation}
Therefore, $v (Y, \tau)$ exists only if $C = - \gamma'_* (h)$. Indeed, multiplying
(\ref{3.2.1}) by $\gamma_*$, integrating the result over $(0, h)$ and then putting
$\tau = \tau_*$, we obtain that
\[ C = \int_0^h \gamma_* \left\{ v'' - [ \tau_*^2 - \omega' (U) ] \, v \right\} \D Y 
= \int_0^h \left( v'' \gamma_* - \gamma_*'' v \right) \D Y ,
\]
because $\gamma_*$ is a normalized Dirichlet eigenfunction of $\D^2 / \D Y^2 +
\omega' (U)$ corresponding to $\tau^2_*$. Now, we integrate by parts in the last
integral and take into account the boundary conditions for $v$ and $\gamma_*$. This
immediately gives that $C = - \gamma'_* (h)$. Then formula (\ref{3.2.0}) follows
from the representation for $\gamma$ with the function $v (Y, \tau)$ found from
problem (\ref{3.2.1}). It is clear that this function exists and is smooth.

\vspace{2mm}

Now we are in a position to prove the proposition about the solvability of the
dispersion equation (\ref{dispers}).

\vspace{2mm}

\noindent {\bf Proposition 3.3.} {\it Let assumptions} (I) {\it and} (II) {\it hold,
and so there exists $s_*$ that solves equation $(\ref{ss2})$ for some ${\cal R}$.
Let $(U, h)$ be the stream solution corresponding to $s_*$. Then the following
assertions are true for the dispersion equation with $\sigma$ defined by $(U, h)$.}

\vspace{2mm}

\noindent (i) {\it If the operator $\D^2 / \D Y^2 + \omega' (U)$ considered on $(0,
h)$ has no Dirichlet eigenvalues, then
\begin{equation}
\sigma (0) = - \frac{3}{2 \kappa} \left[ \frac{\D {\cal R}}{\D s} (s_*) \bigg/
\frac{\D h}{\D s} (s_*) \right] \, . \label{sigma(0)>}
\end{equation}
Moreover, the inequality
\begin{equation}
\frac{\D {\cal R}}{\D s} (s_*) \bigg/ \frac{\D h}{\D s} (s_*) > 0 ,
\label{condRh}
\end{equation}
is a necessary and sufficient condition that equation $(\ref{dispers})$ has one and
only one positive root.}

\vspace{2mm}

\noindent (ii) {\it Let the operator $\D^2 / \D Y^2 + \omega' (U)$ considered on
$(0, h)$ have no zero Dirichlet eigenvalue. If this operator has exactly $k$
positive Dirichlet eigenvalues, then $(\ref{dispers})$ has at least $k$ positive
roots. Moreover, if inequality $(\ref{condRh})$ holds, then $(\ref{dispers})$ has
exactly $k+1$ positive roots.}

\vspace{2mm}

\noindent {\it Proof.} (i) Since the operator $\D^2 / \D Y^2 + \omega' (U)$ has no
Dirichlet eigenvalues, $\sigma$ is a smooth function. In view of Lemma~1.1, we have
to evaluate $\sigma (0)$ in order to prove that (\ref{condRh}) is necessary and
sufficient for the existence of a root. First, we consider the case when $\kappa =
U' (h) > 0$, and show that condition (\ref{condRh}) is equivalent to the following
inequality:
\begin{equation}
\sigma (0) < 0 . \label{0>}
\end{equation}
Then assertion (i) is an immediate consequence of the last two inequalities and the
asymptotic formula (\ref{inf<}), according to which $\sigma (\tau) \to +\infty$ as
$\tau \to +\infty$.

For proving (\ref{sigma(0)>}) and (\ref{0>}) we note that
\begin{equation}
\sigma (0) = \kappa \, \gamma' (h, 0) - \kappa^{-1} + \omega (1) \label{sigma0}
\end{equation}
by the definition of $\sigma$. On the one hand, $\gamma (Y, 0)$ satisfies the
boundary value problem (\ref{gamma}) with $\tau = 0$ and $(U (Y; s_*), h (s_*))$. On
the other hand, the Cauchy problem for a general stream function $U (Y; s)$ is as
follows:
\[ U'' + \omega (U) = 0 , \quad U (0) = 0 , \ \ U' (0) = s . 
\]
Differentiating the first two relations with respect to $s$, we get 
\begin{equation}
\dot{U}'' + \omega' (U) \, \dot{U} = 0 \ \ \mbox{for} \ U \in (0, h) , \quad \dot{U}
(0) = 0 . \label{tau=0}
\end{equation}
(We recall that the top dot denotes the derivative with respect to the parameter
$s$.) Comparing these relations and the problem for $\gamma (Y, 0)$, we see that
\begin{equation}
\gamma (Y, 0) = \frac{\dot{U} (Y; s_*)}{\dot{U} (h (s_*); s_*)} \, , \label{gam0}
\end{equation}
where the denominator does not vanish because $\gamma (Y, 0)$ is well-defined by the
boundary value problem (\ref{gamma}) with $\tau = 0$. Now we differentiate
(\ref{gam0}) and substitute the result into (\ref{sigma0}), where $\kappa$ is
changed to $U' (h (s_*))$ in the first term. Then we get that
\[ \sigma (0) = \frac{U' (h (s_*)) \, \dot{U}' (h (s_*); s_*)}{\dot{U} 
(h (s_*); s_*)} - \frac{1}{\kappa} + \omega (1) = \frac{ \left[ \{U' (h (s))\}^2
\right]_s }{2 \, \dot{U} (h (s); s)} \bigg|_{s=s_*} - \frac{1}{\kappa} + \omega (1)
.
\]
Applying relation (\ref{final}), we obtain
\[ \sigma (0) = \frac{s_* - \omega (h (s_*)) \, \dot{U} 
(h (s_*); s_*)}{\dot{U} (h (s_*); s_*)} - \frac{1}{\kappa} + \omega (1) = -
\frac{1}{\kappa} \left[ 1 + \frac{s_*}{\dot{h} (s_*)} \right] .
\] 
Here $U' (h (s_*))$ is changed back to $\kappa$; moreover the equality $h (s_*) = 1$
and the formula
\[
U' (h (s_*)) \, \dot{h} (s_*) + \dot{U} (h (s_*); s_*) = 0 
\]
are used. The last equality arises when one differentiates the condition $U (h (s);
s) = 1$ (it holds for every stream solution) with respect to $s$ and puts $s=s_*$.
Finally, differentiating ${\cal R}$ with respect to $s$ (see formula (\ref{ss2}),
where ${\cal R}$ the is defined), we get that
\begin{equation}
\sigma (0) = - \frac{1}{\kappa} \left[ 1 + \frac{s_*}{\dot{h} (s_*)} \right] = -
\frac{3}{2 \kappa} \left[ \frac{\D {\cal R}}{\D s} (s_*) \bigg/ \frac{\D h}{\D s}
(s_*) \right] . \label{sigma(0)>!}
\end{equation}
According to this formula, inequality (\ref{0>}) is a consequence of (\ref{condRh})
when $\kappa = U' (h) > 0$, which proves (\ref{0>}). In view of Lemma~1.1, this
completes the proof of the proposition's assertion when the last inequality is
assumed to hold.

\vspace{2mm}

It remains to consider the case when $\kappa = U' (h) < 0$. According to the
asymptotic formula (\ref{inf<}), this condition yields that $\sigma (\tau) \to
-\infty$ as $\tau \to +\infty$. Hence, if we show that inequality (\ref{condRh})
implies that
\begin{equation}
\sigma (0) > 0 , \label{0<}
\end{equation}
then we immediately obtain the required assertion in view of Lemma~1.1. Using
formula (\ref{sigma(0)>!}) (it is independent of the sign of $\kappa$) and the
inequality $\kappa < 0$, we see that (\ref{0<}) is again equivalent to
(\ref{condRh}). Thus the proof of assertion (i) is complete.

\vspace{2mm}

(ii) First, let exactly one Dirichlet eigenvalue $\tau_*^2 > 0$ of $\D^2 / \D Y^2 +
\omega' (U)$ exist. Then formula (\ref{3.2.0}) and the definition of $\sigma$ yield
that $\sigma (\tau) \to \mp \infty$ as $\tau \to \tau_* + 0$ provided $\pm \kappa >
0$. On the other hand, $\sigma (\tau) \to \pm \infty$ as $\tau \to +\infty$ by
virtue of the asymptotic formula (\ref{inf<}). By the assumption there is no
eigenvalue other then $\tau_*^2$. Hence $\sigma (\tau)$ is smooth for $\tau \in
(\tau_*, +\infty)$ and it tends to opposite infinities at the ends of this interval.
Then one and only one root of equation (\ref{dispers}) exists on $(\tau_*, +\infty)$
according to Lemma~1.1.

Since zero is not an eigenvalue, $\sigma (0)$ is defined. Therefore, it is easy to
modify the proof of (i) so that it will combine inequality (\ref{condRh}) and the
limit of $\sigma (\tau)$ as $\tau \to \tau_* - 0$ instead of the limit as $\tau \to
+\infty$. On this way, one obtains that one more root of (\ref{dispers}) exists on
the interval $(0, \tau_*)$.

Now, let us assume that there are exactly two Dirichlet eigenvalues 
\[ \big[ \tau_*^{(1)} \big]^2 \quad \mbox{and} \quad \big[ \tau_*^{(2)} \big]^2 > 
\big[ \tau_*^{(1)} \big]^2 ;
\]
let $\gamma_*^{(1)}$ and $\gamma_*^{(2)}$, respectively, denote the corresponding
eigenfunctions. Then it is easy to see that
\[ \sigma (\tau) = \frac{- \kappa \left[ \big( \gamma_*^{(j)} \big)' (h) 
\right]^2}{2 \tau_*^{(j)} \big( \tau - \tau_*^{(j)} \big)} + O (1) \quad \mbox{as} \
\tau \to \tau_*^{(j)} , \quad j=1,2 ,
\]
which is similar to (\ref{3.2.2}). Thus, $\sigma (\tau)$ tends to opposite
infinities (their signs depend on the sign of $\kappa$) as $\tau$ goes to
$\tau_*^{(1)} + 0$ and  $\tau_*^{(2)} - 0$. Hence, according to Lemma~1.1, there
exists exactly one root of equation (\ref{dispers}) on the interval $\big(
\tau_*^{(1)}, \tau_*^{(2)} \big)$. As in the case of a single eigenvalue, one more
root belongs to $\big( \tau_*^{(2)}, +\infty \big)$, and so the total number of
roots is equal to two. The case of $k$ eigenvalues should be treated in the same
way. The proof is complete.

\vspace{2mm}

\noindent {\bf Remark 3.4.} Note that zero is an eigenvalue of the operator $\D^2 /
\D Y^2 + \omega' (U)$ with the Dirichlet boundary conditions at $Y = 0$ and $Y = h
(s)$ if and only if $\dot{h} (s) = 0$. Indeed, it follows from (\ref{tau=0}) that
$\dot{U} (Y; s)$ is an eigenfunction of this operator corresponding to the zero
eigenvalue, and so $\dot{U} (h (s); s) = 0$. Differentiating the equality $U(h(s),
s) = 0$ with respect to $s$, we get
\[ \dot{h} (s) \, U' (h (s); s) + \dot{U} (h (s); s) = 0 ,
\]
where the second term is equal to zero in view of what is said above. Since $U' (h
(s); s) \neq 0$ by assumption (I), we arrive at the assertion.

\subsection{The dispersion equation for unidirectional flows}

First, a stream solution $(U , h)$ describes a unidirectional flow, that is, $U' (Y)
> 0$ for all $Y \in [0, h]$, only when this solution depends on $s > s_0$ which is a
root of equation $(\ref{ss2})$ with ${\cal R}_0^{(+)} (s)$ (see \cite{KK},
sections~5.2 and 5.3). Moreover, we have (see \cite{KK}, section~5.1):
\begin{equation} 
\pm \frac{\D {\cal R}_0^{(+)}}{\D s} (s) > 0 \quad \mbox{provided} \ \pm (s - s_c) >
0 \, . \label{R0+}
\end{equation}
Second, the corresponding operator $\D^2 / \D Y^2 + \omega' (U)$ considered on $(0,
h)$ has the empty Dirichlet spectrum. Hence the solution $\gamma (Y; \tau)$ of
problem (\ref{gamma}) is defined for all $\tau \in \RR$ and is a smooth function of
both variables. The same is true for $\sigma$ that stands in the dispersion
equation.

\vspace{2mm}

\noindent {\bf Proposition 3.5.} {\it Let $s > s_0$ be a root of the equation ${\cal
R}_0^{(+)} (s) = r$ for some $r > r_c$. If the function $\sigma$ is defined by the
stream solution $(U, h)$ corresponding to this $s$, then the following assertions
hold.}

\noindent (i) {\it For any $s \in (s_0, s_c)$ there exists one and only one positive
solution of equation $(\ref{dispers})$.}

\noindent (ii) {\it Equation $(\ref{dispers})$ has no positive solutions for $s >
s_c$.}

\vspace{2mm}

\noindent {\it Proof.} Assertion (i) (it is included for the sake of completeness)
is an immediate consequence of Lemma~1.1 and Proposition~3.3, (i). The first of
these propositions guarantees the uniqueness of a solution, whereas the second one
yields the existence. Indeed, inequality (\ref{condRh}) follows from (\ref{R0+})
with the lower sign and
\begin{equation}
\frac{\D h_0^{(+)}}{\D s} (s) < 0 . \label{h_0}
\end{equation}
The last inequality holds for all $s > s_0$ according to formula (4.1) in \cite{KK}.

Let us turn to proving (ii). It is shown in the proof of Proposition~3.3 that
equation (\ref{dispers}) has a positive solution for $s > s_c$ if and only if
\[ \frac{\D {\cal R}_0^{(+)}}{\D s} (s) \bigg/ \frac{\D h_0^{(+)}}{\D s} (s) \geq 0 .
\]
However the numerator of the last inequality is positive by (\ref{R0+}), where the
upper sign must be taken, whereas the denominator is negative by (\ref{h_0}). The
obtained contradiction proves assertion (ii).

\section{Proof of Theorem 1.2}

Our proof of Theorem 1.2 is based on the following theorem that deals with
bifurcation from a simple eigenvalue.

\vspace{2mm}

\noindent {\bf Theorem 4.1.} {\it Let $I$ be an open interval of $\RR$ such that $0
\in I$, and let ${\cal X}$, ${\cal Y}$ be Banach spaces. If a continuous map ${\cal
F}: I \times {\cal X} \mapsto {\cal Y}$ has the following properties:}

\vspace{2mm}

{\rm (i)} {\it the equality ${\cal F} (\lambda, 0) = 0$ holds for all $\lambda \in
I$,}

$\!\!${\rm (ii)} {\it the operators ${\cal F}_\lambda$, ${\cal F}_\Phi$ and ${\cal
F}_{\lambda \Phi}$ exist in a neighbourhood of $(0,0)$ and are continuous there,}

$\!\!\!${\rm (iii)} {\it ${\cal F}_\Phi$ is a Fredholm operator with zero index and
the null-space of ${\cal F}_\Phi (0, 0)$ is one-dimensional,}

$\!\!${\rm (iv)} {\it if the null-space of ${\cal F}_\Phi (0, 0)$ is generated by
$\Phi^{(0)}$, then ${\cal F}_{\lambda \Phi} (0, 0) \, \Phi^{(0)}$ does not belong to
the range of ${\cal F}_\Phi (0, 0)$.

\vspace{2mm}

\noindent Then a sufficiently small $\varepsilon > 0$ exists and a continuous curve
\[ \{ (\lambda (t), \, \Phi (t)) : |t| < \varepsilon \} \subset I \times {\cal X} , 
\]
bifurcates from $(0, 0)$. Moreover, for pairs belonging to this curve the following
properties hold:
\[ \Phi (t) = t \, \Phi^{(0)} + o (t) \quad when \ 0 < |t| < \varepsilon ,
\]
and $\{ (\lambda, \, \Phi) \in V : \zeta \neq 0 \ and \ {\cal F} (\lambda, \Phi) =
0\} = \{ (\lambda (t), \, \Phi (t)) : 0 < |t| < \varepsilon \}$, where $V \subset I
\times {\cal X}$ is a certain neighbourhood of $(0, 0)$.}

(v) {\it If ${\cal F}_{\Phi \Phi}$ is also continuous, then the curve is of class
$C^1$.}

\vspace{2mm}

This theorem was proved by Crandall and Rabinowitz (see Theorem~1.7 in \cite{CR}).

\subsection{Application of Theorem 4.1 to equation (\ref{nov4})}

In order to apply Theorem 4.1 to equation (\ref{nov4}) we have to verify conditions
(i)--(iv) of Theorem 4.1 for the operator $F$ assuming that ${\cal X} = {\cal
X}_0^{2, \alpha}$ and ${\cal Y} = {\cal X}^{0, \alpha}$.

\vspace{2mm}

\noindent {\bf Conditions (i)} and {\bf (ii)} immediately follow from the definition
of $F$ [see formulae (\ref{nov4}), (\ref{nov2}) and (\ref{beq'}), (\ref{beq'-})].
Indeed, (\ref{nov2}) and (\ref{beq'}), (\ref{beq'-}) imply that $F (0; \lambda) = 0$
for all $\lambda$, and that the operators $F_\lambda$, $F_\Phi$ and $F_{\lambda
\Phi}$ exist in a neighbourhood of $(0;0)$ and are continuous there.

\vspace{2mm}

\noindent {\bf Condition (iii).} Let us show that $F_\Phi (0; 0) : {\cal X} \mapsto
{\cal Y}$ defined by formulae (\ref{nov5}) and (\ref{nov6}) is a Fredholm operator
with zero index. Introducing the following function (cf. \cite{EEW}, section 3):
\[ \hat \psi = \hat \phi - \frac{z u_z}{h} \hat \zeta \, ,
\]
we get that the operator (\ref{nov5}), (\ref{nov6}) takes the form
\begin{eqnarray} 
&& \ \ \ \ \ \ - \hat \psi_{xx} - \hat \psi_{zz} - \omega' (u) \, \hat \psi \, , 
\label{apr20a} \\ && \left[ \hat \psi_z \right]_{z=h} - \left( \frac{1}{\kappa^2} 
- \frac{\omega (1)}{\kappa} \right) \left[ \hat \psi \right]_{z=h}  \label{apr20b}
\end{eqnarray}
in terms of $\hat \psi$, and the following boundary conditions $\hat \psi (x, 0) =
0$ and $\hat \psi (x, h) = - \kappa \, \hat \zeta (x)$ hold for all $x \in \RR$.

Note that the operator
\[ {\cal X}_0^{2, \alpha} \ni (\hat \phi, \hat \zeta) \mapsto \hat \psi \in \left\{ 
\hat \varphi \in \Pi_{\Lambda_0}^{2, \alpha} (\bar S) : \hat \varphi (x, 0) = 0 \
\mbox{for all} \ x \in \RR \right\}
\]
is an isomorphism. Moreover, the operator, that maps $\hat \psi$ belonging to the
last space into the pair (\ref{apr20a}), (\ref{apr20b}) (this pair belongs to
${\cal X}^{0, \alpha}$), is a  Fredholm operator with zero index. Hence the same is
true for $F_\Phi (0; 0)$.

Furthermore, assumption (II) guarantees that the null-space of $F_\Phi (0; 0)$ is
one-dimension\-al. Indeed, using Proposition 2.1, it is straightforward to check that
this space is generated by the pair $(\phi^{(0)} , \zeta^{(0)})$:
\begin{equation}
\phi^{(0)} (x, z) = w^{(0)} (z) \cos \frac{2 \pi x}{\Lambda_0} \, , \quad
\zeta^{(0)} (x) = \cos \frac{2 \pi x}{\Lambda_0} \, . \label{0_space}
\end{equation}
Here $\Lambda_0$ and $w^{(0)}$ are defined by virtue of a simple positive root
$\tau_0$ of the dispersion equation (\ref{dispers}) for which none of the values
$k \tau_0$ $(k = 1,2,\dots)$ satisfies (\ref{dispers}). Namely, we have that
$\Lambda_0 = 2 \pi / \tau_0$, whereas $w^{(0)}$ satisfies the following relations:
\begin{equation}
- w^{(0)}_{zz} + [ \tau^2_0 - \omega' (u) ] \, w^{(0)} = [ z \, u_z \, \tau^2_0 + 2
\, \omega (u) ] / h \ \ \mbox{on} \ (0, h) , \quad w^{(0)} (0) = w^{(0)} (h) = 0 ,
\label{nov12}
\end{equation}
and
\begin{equation}
w^{(0)}_z (h) = \frac{\kappa}{h} - \frac{1}{\kappa} \, . \label{nov13}
\end{equation}
Note that the last equality is equivalent to $w^{(0)}_z (h) = - \sigma_0 (\tau_0)$
in view of (\ref{nov18}).

\vspace{2mm}

\noindent {\bf Condition (iv).} Let $(\hat \phi, \hat \zeta) \in {\cal X}_0^{2,
\alpha}$, then it is straightforward to check that 
\[ [F_{\lambda\Phi} (0;0)] \, (\hat \phi, \hat \zeta) = \left( \hat \phi_{xx} -
\hat \phi_{zz} - \omega' (u) \hat \phi - \frac{4 \, \omega (u)}{h} \hat \zeta \, ,
\left[ \frac{1}{\kappa} - \frac{\kappa}{h} \right] \! \hat \zeta \right) .
\]
In order to verify condition (iv) we have to show the following. If $(\phi^{(0)} ,
\, \zeta^{(0)})$ belongs to the null-space of $F_\Phi (0;0)$, that is, has the form
described in the previous paragraph, then the range of $F_{\Phi} (0;0)$ does not
contain $[F_{\lambda \Phi} (0;0)] \, (\phi^{(0)} , \, \zeta^{(0)})$. This is a
consequence of the next assertion.

\vspace{2mm}

\noindent {\bf Proposition 4.2.} {\it Let $\tau_0$ be a root of the dispersion
equation $(\ref{dispers}),$ and let assumption {\rm (II)} be fulfilled for $\tau_0$.
Then condition {\rm (iv)} of Theorem~4.1 is equivalent to the relation $\sigma_\tau
(\tau_0) \neq 0$.}

\vspace{2mm}

\noindent {\it Proof.} Note that condition (iv) means that no constant $\varsigma$
and function $W$ exist such that the relations
\begin{eqnarray}
&& - W_{zz} + [ \tau^2_0 - \omega' (u) ] \, W = \frac{\varsigma}{h} [ z \, u_z \,
\tau^2_0 + 2 \, \omega (u) ] \nonumber \\ && \ \ \ - \left[ \tau^2_0 \, w^{(0)} +
w^{(0)}_{zz} + \omega' (u) w^{(0)} + \frac{4 \, \omega (u)}{h} \right] \ \ \mbox{on}
\ (0, h) , \label{nov14} \\ && \ \ \ \ \ \ \ \ \ \ \ \ \ \ \ \ \ \ \ \ W (0) = W (h)
= 0 , \label{nov15} \\ && \ \ \ \ \ \ \ \ W_z (h) = \left( \frac{\kappa}{h} -
\frac{1}{\kappa} \right) \varsigma - \left( \frac{\kappa}{h} - \frac{1}{\kappa}
\right) \label{nov16}
\end{eqnarray}
hold simultaneously. Here $w^{(0)}$ satisfies relations $(\ref{nov12})$ and
$(\ref{nov13})$,

Assuming the contrary, we multiply equation (\ref{nov14}) by $\gamma$, integrate by
parts twice in the left-hand side and use the equation and boundary conditions for
$\gamma$ and the boundary conditions (\ref{nov15}). This yields that
\[ W_z (h) = - \int_0^h \!\! \gamma (t, \tau_0) \left\{ \frac{\varsigma}{h} \left[ 
\tau^2_0 \, t \, u_t + 2 \, \omega (u (t)) \right] - \left[ \tau^2_0 \, w^{(0)} +
w^{(0)}_{tt} + \omega' (u) w^{(0)} + \frac{4 \, \omega (u)}{h} \right] \right\} \D t
.
\]
In view of (\ref{sigma}) we write this as follows:
\[ W_z (h) = - \sigma_0 (\tau_0) \varsigma + \int_0^h \gamma (t, \tau_0) \left[ 
\tau^2_0 \, w^{(0)} + w^{(0)}_{tt} + \omega' (u) w^{(0)} + \frac{4 \, \omega (u)}{h}
\right] \D t \, .
\]
Taking into account formulae (\ref{nov18}) and (\ref{nov16}), it remains to show
that the equality
\[
\int_0^h \gamma (z, \tau_0) \left[ \tau^2_0 \, w^{(0)} + w^{(0)}_{zz} + \omega' (u)
w^{(0)} + \frac{4 \, \omega (u)}{h} \right] \D z = \frac{1}{\kappa} -
\frac{\kappa}{h} 
\]
is not true. The second formula (\ref{sigma}) and the dispersion equation
(\ref{nov18}) imply that the last equality can be written as follows:
\begin{equation}
\int_0^h \gamma (z, \tau_0) \left[ w^{(0)}_{zz} + \tau^2_0 \, w^{(0)} + \omega' (u)
w^{(0)} - \frac{2}{h} \, z \, \tau^2_0 \, u_z \right] \D z = \frac{\kappa}{h} -
\frac{1}{\kappa} \,  \label{nov19}
\end{equation}
Splitting the last integral into the sum of two terms and integrating by parts twice
in the first one, we obtain that the integral is equal to
\begin{eqnarray*}
w^{(0)}_{z} (h) + \int_0^h \left[ \gamma_{zz} (z, \tau_0) + \omega' (u (z)) \,
\gamma (z, \tau_0) \right] \, w^{(0)} (z) \, \D z \\ + \, \tau_0^2 \int_0^h \gamma
(z, \tau_0) \left[ w^{(0)} (z) - \frac{2}{h} z \, u_z (z) \right] \, \D z .
\end{eqnarray*}
In view of equation for $\gamma$ [see (\ref{gamma})], this reduces (\ref{nov19}) to
\begin{equation}
w^{(0)}_{z} (h) + 2 \, \tau_0^2 \int_0^h \gamma (z, \tau_0) \left[ w^{(0)} (z) -
h^{-1} z \, u_z (z) \right] \, \D z = \frac{\kappa}{h} - \frac{1}{\kappa} \, .
\label{nov21}
\end{equation}
Thus, we have to show the impossibility of the latter equality now.

Differentiating formula (\ref{sigma}), we get
\[ \sigma_{0 \tau} (\tau_0) = h^{-1} \int_0^h \left\{ 2 \, \gamma (z, \tau_0) \, 
\tau_0 z u_z (z) + \gamma_\tau (z, \tau_0) [ \tau^2_0 z u_z (z) + 2 \, \omega (u
(z)) ] \right\} \D z .
\]
We also differentiate relations (\ref{gamma}) with respect to $\tau$, thus
obtaining
\[ - \gamma_{\tau zz} + [ \tau^2 - \omega' (u) ] \, \gamma_\tau = - 2 \, \tau \, 
\gamma \ \mbox{on} \ (0, h) , \quad \gamma_\tau (0) = 0 , \ \ \gamma_\tau (h) = 0 .
\]
The equation with $\tau = \tau_0$ we multiply by $w^{(0)}$, integrate over $(0, h)$
and, after integration by parts in the left-hand side, get
\[ \frac{1}{h} \int_0^h \gamma_\tau (z, \tau_ 0) \, \left[ \tau_0 z \, u_z + 2 \,
\omega (u) \right] \D z = - 2 \tau_0 \int_0^h \gamma (z, \tau_0) \, w^{(0)} (z) \,
\D z .
\]
Here problem (\ref{nov18}) is also used. Thus we arrive at
\[ \sigma_{0 \tau} (\tau_0) = 2 \, \tau_0 \int_0^h \gamma (z, \tau_0) \, \left[ 
w^{(0)} (z) - h^{-1} z u_z (z) \right] \D z 
\]
Comparing this, (\ref{nov21}) and (\ref{nov13}), we obtain that $\tau_0 \, \sigma_{0
\tau} (\tau_0) = 0$ which proves our proposition in view of formula (\ref{nov18}).

\vspace{2mm}

Thus condition (iv) is a consequence of Proposition 4.2 and Lemma 1.1.

\vspace{2mm}

\noindent {\bf Condition (v)} follows from formulae (\ref{nov2}), (\ref{beq'}) and
(\ref{beq'-}).

\subsection{Proof of Theorem 1.2}

Since conditions (i)--(v) of Theorem 4.1 are fulfilled for equation (\ref{nov4}),
we have that
\begin{equation}
\phi (x, z, t) = t \, w^{(0)} (z) \cos \frac{2 \pi x}{\Lambda_0} + \phi_* (x, z, t)
, \quad \zeta (x, t) = t \, \cos \frac{2 \pi x}{\Lambda_0} + \zeta_* (x, t) .
\label{jan_1}
\end{equation}
Here $w^{(0)}$ satisfies relations (\ref{nov12}) and (\ref{nov13}), whereas 
\[ \| \phi_* (\cdot , \cdot , t) \|_{\Pi^{2, \alpha}_{\Lambda_0} (\bar S)} = o (t)
\ \ \mbox{and} \ \ \| \zeta_* (\cdot , t) \|_{\Pi^{2, \alpha}_{\Lambda_0}} = o (t) \
\ \mbox{as} \ |t| \to 0 .
\]
It is clear that the second formula (\ref{jan_1}) yields (\ref{xi}).

Substituting the first formula (\ref{jan_1}) into the second formula
(\ref{zeta/phi}) and taking into account (\ref{z}), we obtain that
\[ \psi \left( x, y, t \right) =  u \left( y \frac{h}{\eta (x)} \right) + t \, 
w^{(0)} \left( y \frac{h}{\eta (x)} \right) \cos \frac{2 \pi x}{\Lambda_0} + \phi_*
\left( x, y \frac{h}{\eta (x)}, t \right) .
\]
Using formulae (\ref{nv}), where $\Lambda = \Lambda_0 [1 + \lambda (t)]$, we write
this as follows:
\begin{equation}
\Psi (X, Y, t) = U \left( Y \frac{h}{\xi (X , t)} \right) + t \, w^{(0)} \left( Y
\frac{h}{\xi (X)} \right) \cos \frac{2 \pi X}{\Lambda} + \phi_* \left( \frac{X}{1 +
\lambda (t)} , Y \frac{h}{\xi (X)}, t \right) . \label{last'}
\end{equation}
It is clear that this function is $\Lambda$-periodic and even in $X$. In view of
considerations presented in section~2, $\Psi$ solves problem (\ref{1})--(\ref{3}),
whereas the pair $(\Psi , \xi)$ satisfies condition (\ref{4}). Comparing formulae
(\ref{last'}) and (\ref{estim}), we see that the sum of the last two terms in
(\ref{last'}) is equal to $\Psi_*$ in (\ref{estim}). Combining this fact and the
above mentioned property of $\phi_*$, we get that the second term in the right-hand
side of (\ref{estim}) is $O (t)$. This completes the proof of Theorem 1.2.

\section{Operator form of problem (\ref{1})--(\ref{4}) \\ with fixed wavelength}

Here we reduce problem (\ref{1})--(\ref{4}) to an operator equation with fixed
wavelength and verify conditions (i)--(iv) of the Crandall--Rabinowitz theorem for
this equation which proves Theorem 1.3.

\subsection{Derivation of an operator equation}

For reducing problem (\ref{1})--(\ref{4}) to an operator equation with the parameter
$\mu$, we begin with reformulating the problem for $U (Y; s)$. Changing $Y \in (0,
h (s))$ to $z = Y h / h (s) \in (0, h)$, we put
\[ u^* (z; s) = U \left( z \frac{h (s)}{h}; s \right) .
\]
Then $u^*$ satisfies the following relations:
\begin{eqnarray}
&& u^*_{zz} + \left[ \frac{h (s)}{h} \right]^2 \omega (u^*) = 0 \ \ \mbox{on} \ (0,
h) ; \quad u^* (0; s) = 0 , \ \ u^* (h; s) = 1 ; \label{dec1} \\ && \ \ \ \ \ \ \ \
\ \ \ \ \ \ \ \ \ \ \ \ \ \ \left[ \frac{h}{h (s)} \, u^*_{z} (h) \right]^2 = 3 \, r
(s) - 2 \, h (s) . \label{dec2}
\end{eqnarray}
To reformulate problem (\ref{1})--(\ref{4}) with $r (s)$ instead of $r$, we put
$x=X$, change $Y \in (0, \xi (x))$ to
\[ z = Y  \frac{h}{\xi (x)} \in (0, h) \quad \mbox{and introduce} \quad
\Psi^* (x, z) = \Psi \left( x, \frac{z}{h} \, \xi (x) \right) .
\]
The last function defined on $\bar S = \RR \times [0, h]$, required to be
$\Lambda_0$-periodic and even in $x$, and to satisfy the boundary conditions
\[ \Psi^* (x, 0) = 0 \ \ \mbox{and} \ \ \Psi^* (x, h) = 1 \ \ \mbox{for all} \
x \in \RR ,
\] 
we subject to the following weak setting [cf. formula (\ref{iiphi'})]:
\begin{equation}
\int_{-\Lambda_0/2}^{\Lambda_0/2} \int_0^h \bigg\{ \left[ \Psi^*_x - \frac{z
\xi_x}{\xi} \Psi^*_z \right] \left[ v_x - \frac{z \xi_x}{\xi} v_z \right] + \left(
\frac{h}{\xi} \right)^2 \Psi^*_z \, v_z - \omega (\Psi^*) \, v \bigg\} \frac{\xi \,
\D x \, \D z}{h} = 0 . \label{dec3}
\end{equation}
This integral identity must hold for all $v \in W^{1,2} (S)$ that are
$\Lambda_0$-periodic in $x$ and vanish at $z=0$ and $z = h$. Bernoulli's equation
for $\Psi^*$ and $\xi$ takes the form:
\begin{equation} 
\Psi_z^* (x, h) - \frac{\xi (x)}{h} \left[ \frac{3 \, r (s) - 2 \, \xi
(x)}{1+\xi_x^2 (x)} \right]^{1/2} = 0 \quad \mbox{for all} \ x \in \RR .
\label{dec4}
\end{equation}
As in section 2.2, we define new unknown functions $\zeta^* (x)$ and $\phi^* (x, z)$
on $\RR$ and $\bar S$, respectively, by virtue of the following relations:
\begin{equation}
\xi (x) = h + \zeta^* (x) \quad \mbox{and} \quad \Psi^* (x, z) = u^* (z) + \phi^*
(x, z) , \label{dec5}
\end{equation}
and so $\zeta^*$ and $\phi^*$ are small for a perturbed flow bifurcating from $(U,
h)$. Besides, these functions are $\Lambda_0$-periodic, even in $x$ and satisfy the
boundary conditions [they follow from (\ref{dec1})]:
\begin{equation}
\phi^* (x, 0) = \phi^* (x, h) = 0 \ \ \mbox{for all} \ x \in \RR . \label{dec6}
\end{equation}

Now we are in a position to write down an operator equation for $\Phi^* = (\phi^*,
\zeta^*)$ in the same way as in section 2.2. First, using (\ref{dec5}) and
(\ref{dec6}), we obtain from (\ref{dec3}) 
\begin{eqnarray}
&& \ \ \ \ \ \ \ \ \ \ \ \ \ \ \ \ \ \ \ \ \ \ \ \ \ \ \ \ \ \ \ \ \ F_1^* (\Phi^*;
\mu) = 0 , \quad \mbox{where} \label{dec7} \\ && F_1^* (\Phi^*; \mu) = - \bigg\{
\frac{\xi}{h} \left[ \phi_x^* - \frac{z \zeta_x^*}{\xi} (u_z^* + \phi_z^*) \right]
\bigg\}_x + \bigg\{ \frac{z \zeta_x^*}{h} \left[ \phi_x -\frac{z \zeta_x^*}{\xi}
(u_z^* + \phi_z^*) \right] \bigg\}_z \nonumber \\ && \ \ \ \ \ \ \ \ \ \ \ \ \ \ \ \
\ \ \ \ \ \ \ \ \ \ \ \ - \frac{h}{\xi} \, (u^* + \phi^*)_{zz} - \frac{\xi}{h}
\, \omega (u^* + \phi^*) \, . \nonumber
\end{eqnarray}
The first relation (\ref{dec1}) allows us to transform the last line
\begin{eqnarray}
&& F_1^* (\Phi^*; \mu)= - \bigg\{ \frac{\xi}{h} \left[ \phi_x^* - \frac{z
\zeta_x^*}{\xi} (u_z^* + \phi_z^*) \right] \bigg\}_x + \bigg\{ \frac{z \zeta_x^*}
{h} \left[ \phi_x -\frac{z \zeta_x^*}{\xi} (u_z^* + \phi_z^*) \right] \bigg\}_z
\nonumber \\ && \ \ \ \ \ \ \ \ \ \ \ \ \ \ \ \ \, - \, \frac{h}{\xi} \, \phi^*_{zz}
- \frac{\xi}{h} \left[ \omega (u^* + \phi^*) - \omega (u^*) \right] - \frac{2 h
\zeta^* + \zeta^{*2}}{h \xi} \, \omega (u^*) . \label{dec8}
\end{eqnarray}
Second, (\ref{dec4}) and (\ref{dec5}) imply
\begin{eqnarray}
&& \ \ \ \ \ \ \ \ \ \ \ \ \ \ \ \ \ \ \ F_2^* (\Phi^*; \mu) = 0 , \quad
\mbox{where} \label{dec9} \\ && \ \ \ \ F_2^* (\Phi^*; \mu) = \phi_z^* -
\frac{\xi}{h} \left[ \frac{3 \, r^* (\mu) - 2 \, \xi}{1 + \zeta_x^2} \right]^{1/2} +
u^*_z (h) \nonumber \\ && = \phi_z^* - \frac{\xi}{h} \left[ \frac{3 \, r^* (\mu) - 2
\, \xi}{1 + \zeta_x^{*2}} \right]^{1/2} + \frac{h^* (\mu)}{h} \left[ 3 \, r^*
(\mu) - 2 \, \xi \right]^{1/2} . \label{dec10}
\end{eqnarray}
The last equality is a consequence of (\ref{dec2}); here and below $r^* (\mu)$ and
$h^* (\mu)$ stand for $r (s_* + \mu)$ and $h (s_* + \mu)$, respectively.

Combining (\ref{dec7}) and (\ref{dec9}), we write problem (\ref{1})--(\ref{4}) as
the following operator equation:
\begin{equation}
F^* (\Phi^*; \mu) = 0 , \quad \mbox{where} \ F^* = (F_1^*, F_2^*)  \label{dec11}
\end{equation}
is the nonlinear operator whose components $F_1^*$ and $F_2^*$ are given by formulae
(\ref{dec8}) and (\ref{dec10}), respectively. We see that $F^*$ maps ${\cal X}_0^{2,
\alpha} \times (-\delta, \delta) \mapsto {\cal X}^{0, \alpha}$ continuously; here
$\delta$ is a sufficiently small positive number.

\subsection{Application of Theorem~4.1 to equation (\ref{dec11})}

{\bf Condition (i).} Formula (\ref{dec8}) immediately yields that $F_1^* (0; \mu)$
vanishes for all $\mu \in (-\delta, \delta)$, and so it remains to check the same
for $F_2^* (0; \mu)$. For this purpose we note that the second expression in formula
(\ref{dec10}) can be written as follows:
\begin{eqnarray}
&& F_2^* (\Phi^*; \mu) = \phi_z^* + \frac{2 \, h^* (\mu) \, \zeta^*}{h \left\{
\left[ 3 \, r^* (\mu) - 2 \, \xi \right]^{1/2} + \left[ 3 \, r^* (\mu) - 2 \, h^*
(\mu) \right]^{1/2} \right\}} \nonumber \\ && \ \ \ \ \ \ \ \ \ \ \ \ \ \, - \left[
3 \, r^* (\mu) - 2 \, \xi \right]^{1/2} \left[ \frac{\zeta^*}{h} + \frac{\xi}{h}
\left( 1 - \frac{1}{\sqrt{1 + \zeta^{*2}_x}} \right) \right] , \label{dec14}
\end{eqnarray}
which obviously vanishes for $\Phi^* = 0$ and all $\mu \in (-\delta, \delta)$.

\vspace{2mm}

\noindent {\bf Condition (ii)} that the operators $F_\mu^*$, $F_{\Phi^*}^*$ and $F_{\mu
\Phi^*}^*$ exist in a neighbourhood of $(0;0)$ and are continuous there is an
immediate consequence of formulae (\ref{dec8}) and (\ref{dec10}).

\vspace{2mm}

\noindent {\bf Condition (iii).} It immediately follows from formulae (\ref{dec8})
and (\ref{dec14}) that the components of $F_{\Phi^*}^* (0;0)$ are given by
(\ref{nov5}) and (\ref{nov6}) with $u$ changed to $u^*$. Hence the considerations
applied in section 4.1 for verifying condition (iii) remain valid in the present
case, and so the one-dimensional null-space of $F_{\Phi^*}^* (0;0)$ is generated by
the pair $(\phi^{(0)} , \zeta^{(0)})$ defined by formulae
(\ref{0_space})--(\ref{nov13}).

\vspace{2mm}

\noindent {\bf Condition (iv).} It is straightforward to find the Fr\'echet
derivative $F_{\Phi^*}^* (0;\mu)$. Indeed, formulae (\ref{dec8}) and (\ref{dec14})
yield that this operator being applied to $(\hat \phi, \hat \zeta) \in {\cal
X}_0^{2, \alpha}$ has the following components:
\begin{eqnarray}
&& - \frac{h^* (\mu)}{h} \hat \phi_{xx} - \frac{h}{h^* (\mu)} \hat \phi_{zz} -
\frac{h^* (\mu) \, \omega' (u^*)}{h} \hat \phi + \frac{z \, u_z^*}{h} \hat
\zeta_{xx} - \frac{2 \, \omega (u^*)}{h} \hat \zeta \, , \label{dec12} \\ && \ \ \ \
\ \ \ \ \ \ \ \ \ \ \ \ \ \ \ \left[ \hat \phi_z \right]_{z=h} - \left[
\frac{\kappa^* (\mu)}{h} - \frac{h^* (\mu)}{h \kappa^* (\mu)} \right] \hat \zeta  ,
\label{dec13}
\end{eqnarray}
where $\kappa^* (\mu) = \left[ 3 \, r^* (\mu) - 2 \, h^* (\mu) \right]^{1/2}$.

\vspace{2mm}

\noindent {\bf Proposition 5.1.} {\it Let $\tau_0$ be a root of the dispersion
equation $(\ref{dispers_*})$ with $s = s_*$, and let assumption {\rm (II)} be
fulfilled for $\tau_0$. Then condition {\rm (iv)} of Theorem~4.1 is valid if and
only if the relation  $\dot{\sigma}_* (\tau (s_*); s_*) \neq 0$ holds.}

\vspace{2mm}

We recall that the top dot denotes the derivative with respect to the parameter
$s$.

\vspace{2mm}

\noindent {\it Proof.} Let a pair $(V (z), \rho) \cos \tau_0 x$ with constant
$\rho$ belong to the kernel of $F_{\Phi^*}^* (0;\mu)$, that is, the following
relations hold for $\tau = \tau_0$ [cf. (\ref{dec12}) and (\ref{dec13})]:
\begin{eqnarray}
&& - V_{zz} + \left[ \frac{h^* (\mu)}{h} \right]^2 \left[ \tau^2 - \omega' (u^*)
\right] V = \frac{h^* (\mu) \rho}{h^2} \left[ z \, u_z^* \, \tau^2 + 2 \, \omega
(u^*) \right] \ \ \mbox{on} \ (0, h) , \label{dec15} \\ && \ \ \ \ \ \ \ \ \ \ \ \ \
\ \ \ \ \ \ \ \ \ \ \ \ \ \ \ \ \ \ \ \ \ \ \ \ \ V (0) = V (h) = 0 ; \label{dec16}
\\ && \ \ \ \ \ \ \ \ \ \ \ \ \ \ \ \ \ \ \ \ \ \ \ \ \ \ \ \ V_z (h) - \left[
\frac{\kappa^* (\mu)}{h} - \frac{h^* (\mu)}{h \kappa^* (\mu)} \right] \rho = 0 .
\label{dec17}
\end{eqnarray}
Note that the function $\gamma^{(0)} (z) = \gamma_* (z \, h (s) / h , \tau_0; s)$
(see (\ref{gamma_*}) for the definition of $\gamma_*$) satisfies the following
problem:
\begin{equation}
- \gamma^{(0)}_{zz} + \left[ \frac{h (s)}{h} \right]^2 \left[ \tau^2_0 - \omega'
(u_*) \right] \, \gamma^{(0)} = 0 \ \mbox{on} \ (0, h), \quad \gamma^{(0)} (0) = 0 ,
\ \ \gamma^{(0)} (h) = 1 . \label{gamma_0}
\end{equation}
We multiply equation (\ref{dec15}) by $\gamma^{(0)}$ and integrate the result over
$(0, h)$. Then integration by parts yields that
\[ V_z (h) = - \frac{h^* (\mu) \rho}{h^2} \int_0^h \gamma^{(0)} (z) \left[ z \, 
u_z^* \, \tau^2_0 + 2 \, \omega (u^*) \right] \D z .
\]
Then the first relation (\ref{dec1}), its consequence
\[ \left[ \omega (u^*) \right]_z = \omega' (u^*) \, u^*_z = - \left[
\frac{h^* (\mu)}{h} \right]^2 u^*_{zzz}
\]
and problem (\ref{gamma_0}) yield that
\begin{eqnarray*}
&& V_z (h) = - \frac{h^* (\mu) \rho}{h^2} \int_0^h \left\{ \left( \left[
\frac{h}{h^* (\mu)} \right]^2 \gamma^{(0)}_{zz} + \omega' (u^*) \gamma^{(0)} \right)
z \, u_z^* - 2 \left[ \frac{h}{h^* (\mu)} \right]^2 \gamma^{(0)} u^*_{zz} \right\}
\D z \\ && \ \ \ \ \ \ \ \ \, = - \frac{\rho} {h^* (\mu)} \int_0^h \left[
\gamma^{(0)}_{zz} z \, u_z^* - z \gamma^{(0)} u_{zzz}^* - 2 \gamma^{(0)} u^*_{zz}
\right] \D z .
\end{eqnarray*}
Integrating by parts, we get
\begin{eqnarray*}
&& V_z (h) = - \frac{\rho} {h^* (\mu)} \left\{ - h \, u_{zz}^* (h) + \int_0^h \left[
\gamma^{(0)}_{zz} z \, u_z^* - z \gamma^{(0)}_z u_{zz}^* - \gamma^{(0)} u^*_{zz}
\right] \D z \right\} \\ && \ \ \ \ \ \ \ \ \, = - \frac{\rho} {h^* (\mu)} \left[ -
h \, u_{zz}^* (h) + h \gamma^{(0)}_z (h) u_z^* (h) - u_z^* (h) \right] .
\end{eqnarray*}
Now, relations (\ref{dec1}), (\ref{dec2}) and the definition of $\kappa$ reduce this
to
\[ V_z (h) = - \rho \left[ \frac{h^* (\mu)}{h} \omega (1) + \kappa^* (\mu) 
\gamma^{(0)}_z (h) - \frac{\kappa^* (\mu)}{h} \right] .
\]
Combining this and (\ref{dec17}), we arrive at
\begin{equation}
\kappa^* (\mu) \gamma^{(0)}_z (h) = \frac{h^* (\mu)}{h \kappa^* (\mu)} - \frac{h^*
(\mu)}{h} \omega (1) , \label{dec18}
\end{equation}
which is a necessary condition for $(V, \rho)$ to satisfy
(\ref{dec15})--(\ref{dec17}). In view of Lemma 2.2, the latter problem is solvable
for $h (s)$ close to $h$.

Now we are in a position to complete our proof. Let us differentiate the relation
$[F_{\Phi^*}^* (0;\mu)] (V, \rho) = 0$ (see (\ref{dec15})--(\ref{dec17}) for its
components) with respect to $\mu$ or, what is the same, with respect to $s$, thus
obtaining for $s = s_*$ (that is, $\mu = 0$):
\[ [F_{\Phi^*}^* (0;0)] (V_s, 0) + [F_{s \Phi^*}^* (0;0)] (V, \rho) + 2 \tau
\dot{\tau} (s_*) \left( V - \frac{z}{h} u^*_z \right) = 0
\]
and
\[ V_{s z} (h) - \left[ \left\{ \frac{\kappa (s)}{h} - \frac{h (s)}
{h \kappa (s)} \right\}_s \, \right]_{s=s_*} \!\! \rho = 0 .
\]
As in the proof of Proposition 4.2, it is sufficient to show that there are no
constant $\varsigma^*$ and function $W^* (z)$ such that the following relations
\begin{eqnarray}
&& [F_{\Phi^*}^* (0;0)] (W^*, 0) + 2 \tau_0 \dot{\tau} (s_*) \left( W^* - \frac{z}{h}
u^*_z \right) = 0 , \label{dec19} \\ && \ \ \ \ \ \ \ \ \ \ \ W_z^* (h) - \left[
\frac{\kappa}{h} - \frac{1}{\kappa} \right] \varsigma^* = 0 \nonumber
\end{eqnarray}
hold simultaneously.

In the same way as in the proof of Proposition 2.1, it follows from (\ref{dec19})
that
\[ W^*_z (h) = 2 \tau_0 \dot{\tau} (s_*) \int_0^h \gamma (z,\tau) \left[ W^* (z) - 
\frac{z}{h} u^*_z \right] \D z \, ,
\]
which gives [cf. formulae (\ref{sigma})]:
\begin{eqnarray*}
&& \ \ \ \ W^*_z (h) = \dot{\tau} (s_*) \, \dot{\sigma}_0^{*} (\tau (s_*)) , \quad
\mbox{where} \\ && \sigma_0^{*} (\tau (s)) = \kappa (s) \, \gamma^{(0)} (h) -
\frac{h (s)} {h} \left[ \frac{1}{\kappa (s)} - \omega (1) \right] .
\end{eqnarray*}
Using this and (\ref{dec18}), one arrives at the proposition's assertion.

\vspace{2mm}

Thus condition (iv) is valid in view of Proposition 5.1 and assumption (III).

\vspace{2mm}

\noindent {\bf Condition (v)} follows from formulae (\ref{dec8}) and (\ref{dec10}).

\vspace{2mm}

Since conditions (i)--(v) of Theorem~4.1 hold for equation (\ref{dec11}) under
assumptions (I)--(III), the same considerations as in section~4.2 prove Theorem 1.3.

\section{Examples}

\begin{figure}[t]
 \vspace{3mm}
\begin{center}
  \SetLabels
  \L (-0.14*0.89) ${\cal R}_0^{(+)}, {\cal R}_1^{(+)}$ \\
  \L (0.85*0) $s$\\
  \endSetLabels
  \leavevmode\AffixLabels{\includegraphics[width=90mm]{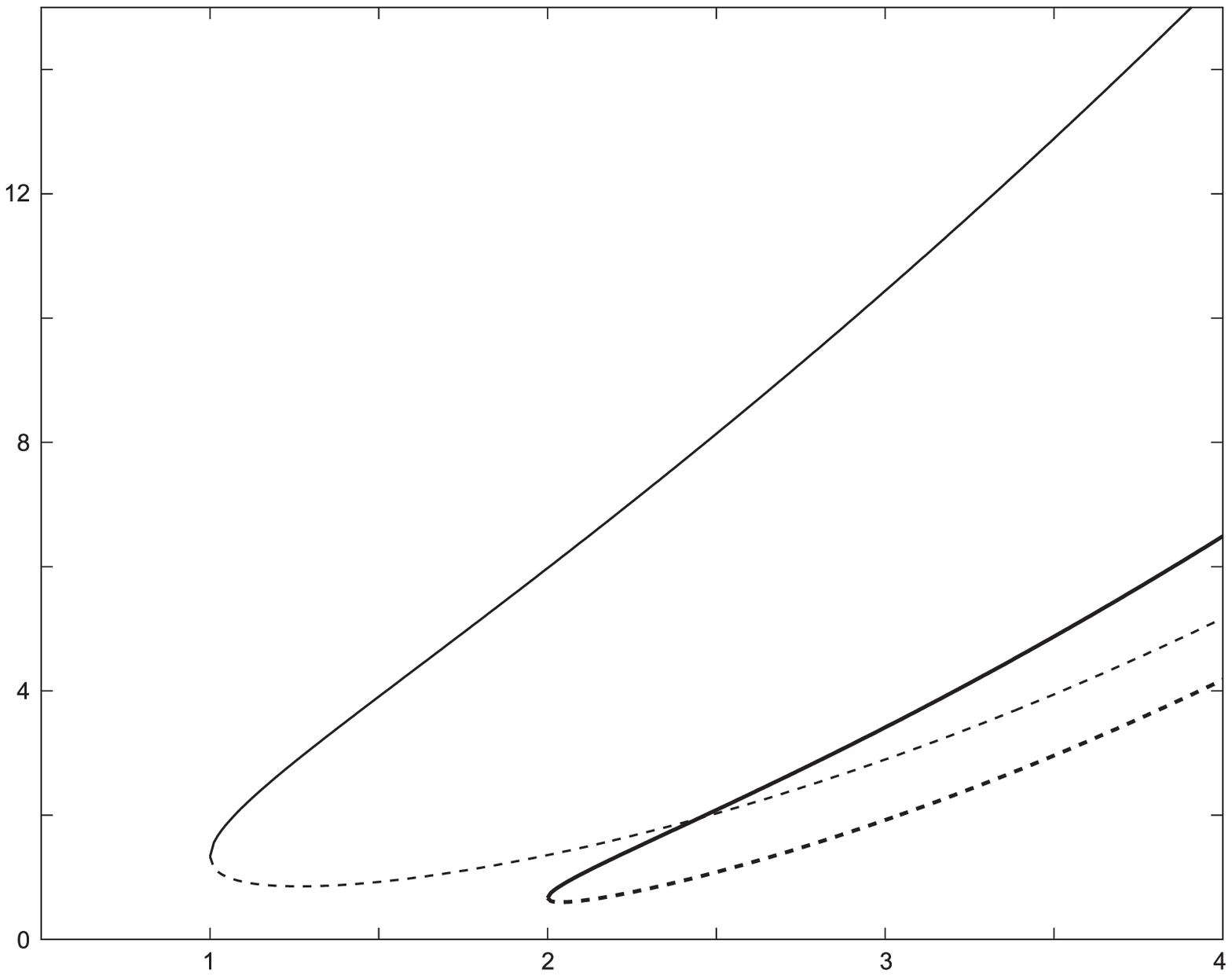}}
  \end{center}
  \vspace{-3.5mm}
  \caption{For the constant vorticity $\omega = b > 0$, the functions 
  ${\cal R}_0^{(+)}$ (dashed lines) and ${\cal R}_1^{(+)}$ (bold lines) 
  are plotted for $b=1/2$ (left) and $b=2$ (right).}
\end{figure}

In this section, we consider two examples of vorticity distributions (the
corresponding flows of constant depth were studied in our paper \cite{KK}), for
which the dispersion equation is investigated along with condition (\ref{condRh}).
This allows us to draw conclusions about the existence of Stokes waves perturbing a
flow of constant depth with vorticity. In the first example, the vorticity is equal
to a positive constant, while the second example deals with the linear vorticity
having a positive coefficient.

\subsection{Constant positive vorticity}

Let $\omega = b$ be a positive constant, then it immediately follows (see \cite{KK}
section~6.1) that $s_0 = \sqrt{2 b}$, and
\begin{equation}
h^{(+)}_0 (s) = \frac{s - \sqrt{s^2 - 2 b}}{b} \, , \quad h^{(+)}_1 (s) = \frac{s +
\sqrt{s^2 - 2 b}}{b} \label{N2}
\end{equation}
are the only non-vanishing depth functions defined for $s \geq s_0$ and such that
$h^{(+)}_0 (s) \leq h^{(+)}_1 (s)$ with the equality holding only for $s = s_0$ when
both values are equal to $h_0 = \sqrt{2 / b}$. Hence the left-hand sides in the
Bernoulli's equation (\ref{ss2}) are as follows (see Fig.~1, where these functions
are plotted for $b=1/2$ and $b=2$; reproduced from \cite{KK}, p.~391):
\begin{equation}
{\cal R}_{\frac{1}{2} \mp \frac{1}{2}}^{(+)} (s) = \frac{1}{3} \left[ s^2 - 2 b + 2
\frac{s \mp \sqrt{s^2 - 2 b}}{b} \right] . \label{feb14}
\end{equation}
These functions have the following properties: ${\cal R}_0^{(+)} (s) < {\cal
R}_1^{(+)} (s)$ for $s > s_0$, while
\[ r_0 = {\cal R}_0^{(+)} (s_0) = {\cal R}_1^{(+)} (s_0) = \frac{2}{3}
\sqrt{\frac{2}{b}} < +\infty ,
\]
and
\begin{equation}
{\cal R}_{\frac{1}{2} \mp \frac{1}{2}}^{(+)} (s) = \frac{s^2}{3} + \frac{2}{3 b} (s
\mp s) + O (1) \quad {\rm as}\ s \to +\infty . \label{12mp12}
\end{equation}
The distance between ${\cal R}_0^{(+)} (s)$ and ${\cal R}_1^{(+)} (s)$ increases
with $s$ due to the second term in the right-hand side. [This term is lost in
formula (6.2), \cite{KK}, analogous to (\ref{12mp12}).] Furthermore, we have that
\begin{equation}
\frac{\D {\cal R}_{\frac{1}{2} \mp \frac{1}{2}}^{(+)}}{{\D} s} (s) \to \mp \infty
\quad {\rm as}\ s \to s_0 + 0 \quad {\rm and} \quad \frac{\D {\cal R}_1^{(+)}}{{\D}
s} (s) > 0 \quad {\rm for}\ s > s_0 . \label{R_s-}
\end{equation}
The first of these relations shows that the graphs of ${\cal R}_0^{(+)}$ and ${\cal
R}_1^{(+)}$ have the common vertical tangent at the point $(s_0, r_0)$, while the
latter inequality implies that the function ${\cal R}_1^{(+)}$ increases strictly
monotonically, and so its graph of lies strictly above the level $r_0$. Finally,
the single minimum $r_c < r_0$ of ${\cal R}_0^{(+)}$ is attained at $s_c > s_0$
satisfying the following equation:
\begin{equation}
(s_c b)^2 - 2 b^3 = \left( \frac{s_c b}{1 + s_c b} \right)^2 . \label{root1}
\end{equation}

It is clear that the first component of the stream solution is $U (Y; s) = -
\frac{b}{2} Y^2 + s Y$. It solves the corresponding Cauchy problem (see
(\ref{cauchy}) for the Cauchy data). Since there are two options for the second
component [they are given by formulae (\ref{N2})], two steady flows of constant
depth exist for every $r > r_c$. It is shown in \cite{KK}, p.~392, that if the depth
is equal to $h^{(+)}_0 (s)$ and $s \in (s_0, s_c)$ is obtained from equation
(\ref{ss2}) with ${\cal R}_0^{(+)}$ [it will be clear from what follows why $s >
s_c$ is not considered], then the flow is unidirectional because $U' (Y; s) > 0$ for
all $Y \in [0, \, h^{(+)}_0 (s)]$. On the contrary, if the depth is equal to
$h^{(+)}_1 (s)$ [in this case, $s > s_0$ satisfies equation (\ref{ss2}) with ${\cal
R}_1^{(+)}$], then there is a near-surface counter-current. Indeed, it is easy to
calculate that
\[ U' (h^{(+)}_1 (s); s) = - \sqrt{s^2 - 2b} < 0 \quad \mbox{and} \quad U' (Y; s) < 0
\ \mbox{for} \ Y > s/b
\]
It is clear that the latter value is less than $h^{(+)}_1 (s)$.

\vspace{2mm}

In order to write down the dispersion equation we find the following solution of
problem (\ref{gamma}):
\[ \gamma (Y, \tau) = \frac{\sinh \tau Y}{\sinh \tau h} , \quad Y \in [0, h (s)] \, .
\]
Here and below $h = h (s)$ is either $h^{(+)}_0 (s)$ or $h^{(+)}_1 (s)$ [see
(\ref{N2})], and $s$ is a root of the corresponding equation (\ref{ss2}); the common
form of both these equations is as follows:
\[ s^2 + 2 h (s) = 3 r + 2 b \, . 
\]
We substitute $\kappa = -(b h - s)$ and $\gamma$ into (\ref{dispers}) and obtain
\begin{equation}
\tau (b h - s) \coth \tau h - b - (b h - s)^{-1} = 0 . \label{dispers_b}
\end{equation}
This dispersion equation coincides with (\ref{coth}) when $b=0$. Indeed, we have that
$s = h = h^{(+)}_0$ in view of the first formula (\ref{N2}) [the second formula
(\ref{N2}) is meaningless for $b=0$].

It is easy to see that the derivative of the left-hand side in (\ref{dispers_b}) has
a definite sign that depends on the sign of $b h - s$, and so if this equation has a
root, then it is simple. Instead of demonstrating directly that (\ref{dispers_b}) is
solvable for both $h^{(+)}_0 (s)$ and $h^{(+)}_1 (s)$ [see (\ref{N2})], we apply
Proposition~3.3 for this purpose.

First, let $r > r_0$, then the second relation (\ref{R_s-}) implies that the
function ${\cal R}_1^{(+)}$ increases monotonically. Thus, the inequality
\[ \frac{\D h_1^{(+)}}{\D s} (s) = \frac{1}{b} \left( 1 + \frac{s}{\sqrt{s^2 - 2b}}
\right) > 0
\]
guarantees that condition (\ref{condRh}) is fulfilled, and so Proposition~3.3 yields
that (\ref{dispers_b}) with $h = h^{(+)}_1 (s)$ has a positive root. In this case,
the left-hand side in (\ref{dispers_b}) is a monotonically increasing function of
$\tau$ because $b h - s = \sqrt{s^2 - 2b} > 0$. Hence the dispersion equation
(\ref{dispers_b}) with $h = h^{(+)}_1 (s)$ has only one positive root.

Let us turn to the case when $h = h^{(+)}_0 (s)$ and $s \in (s_0, s_c)$. Then we
have that $r \in (r_c, r_0)$ and
\[ \frac{\D h_0^{(+)}}{\D s} (s) = \frac{1}{b} \left( 1 - \frac{s}{\sqrt{s^2 - 2b}}
\right) < 0 \quad \mbox{for all} \ s > s_0 .
\]
Moreover, in view of (\ref{root1}) the following inequality holds
\[ \frac{\D {\cal R}_0^{(+)}}{\D s} (s) = \frac{2}{3} \left[ s + \frac{1}{b} 
\left( 1 - \frac{s}{\sqrt{s^2 - 2b}} \right) \right] < 0 \quad \mbox{for} \ s \in
(s_0, s_c) .
\]
Hence Proposition~3.3 yields that the dispersion equation (\ref{dispers_b}) with $h
= h^{(+)}_0 (s)$ has a positive root. Again, there is only one root because the
left-hand side in (\ref{dispers_b}) is a monotonically decreasing function of
$\tau$, since $b h - s = -\sqrt{s^2 - 2b} < 0$ in the present case.

Combining the above considerations and Theorem 1.2, we arrive at the following.

\vspace{2mm}

\noindent {\bf Proposition 6.1.} {\it For every $s \in (s_0, s_c)$ there exists a
brunch of flows having small-amplitude Stokes waves on their free surfaces; for all
these flows $r = {\cal R}_0^{(+)} (s)$ $($see $(\ref{feb14})$ with the minus
sign$)$. This brunch bifurcates from the horizontal shear flow whose depth
$h_0^{(+)} (s)$ is given by the first formula $(\ref{N2})$.

For every $s \in (s_0, +\infty)$ there exists an analogous brunch of flows for which
$r = {\cal R}_1^{(+)} (s)$ $($see $(\ref{feb14})$ with the plus sign$)$. This brunch
bifurcates from the horizontal shear flow whose depth $h_1^{(+)} (s)$ is given by
the second formula $(\ref{N2})$.}

\vspace{2mm} 

\noindent {\bf Remark 6.2.} Further details about small-amplitude Stokes waves with
constant vorticity can be found in \cite{Wah} and \cite{CV}. In the paper
\cite{Wah}, Wahl\'en paid much attention to the behaviour of streamlines for such a
flow. In particular, he investigated in detail the streamlines that form a `critical
layer' which separates two layers with the opposite directions of flow. Moreover,
within this layer, all streamlines are closed, that is, it consists of the so-called
cat's-eye vortices (see figures~1 and 3 in \cite{Wah}, where $\omega = b < 0$).
Besides, a full description of particle paths is given in \cite{Wah}. The approach
developed for the same problem by Constantin and Varvaruca \cite{CV} is applicable
to a wider set of free surface profiles, in particular, it includes overhanging
ones.

\begin{figure}[t]
 \vspace{3mm}
\begin{center}
  \SetLabels
  \L (-0.04*0.89) ${\cal R}_j^{(\pm)}$\\
  \L (0.85*-0.01) $s$\\
  \endSetLabels
  \leavevmode\AffixLabels{\includegraphics[width=90mm]{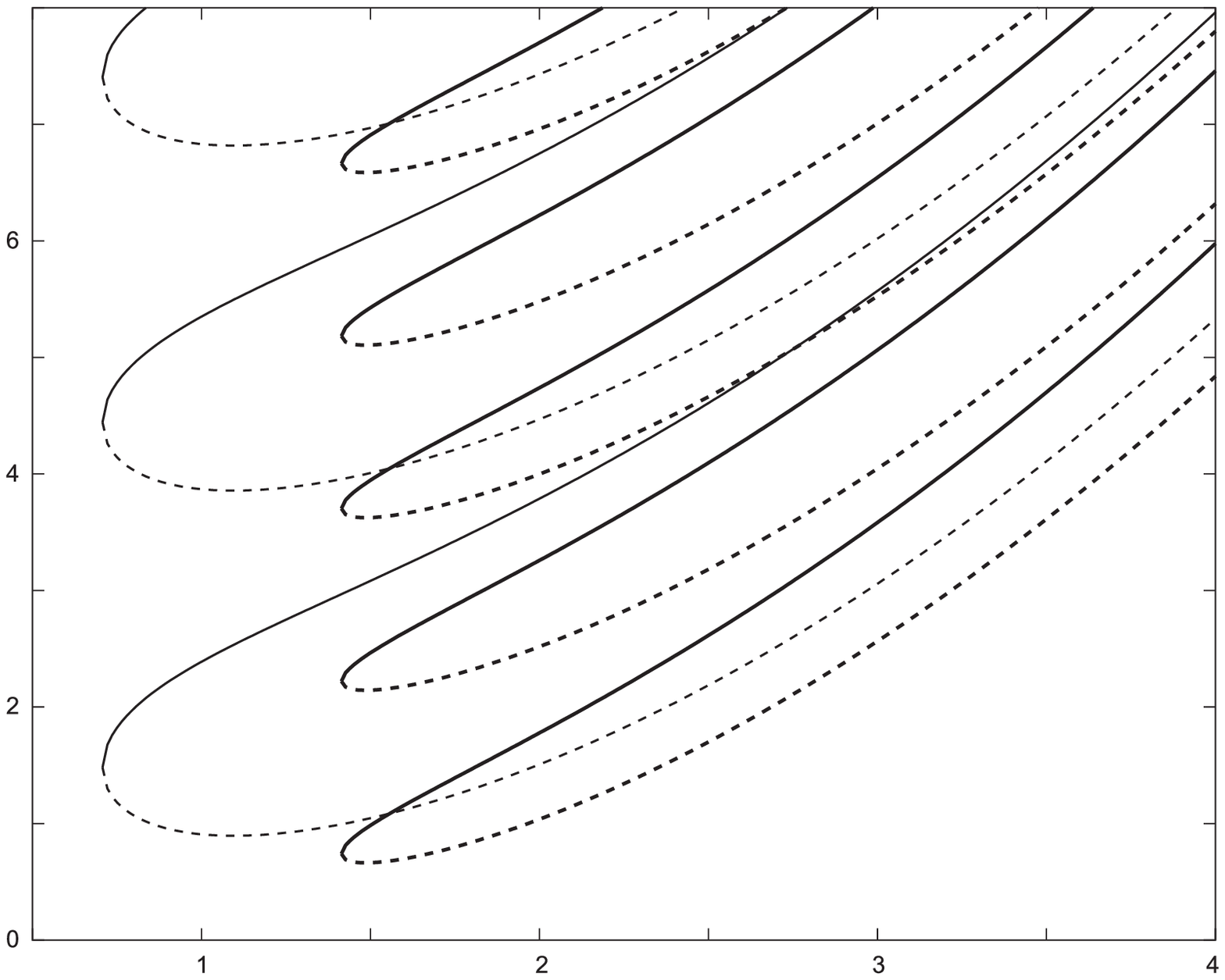}}
  \end{center}
  \vspace{-3.5mm}
  \caption{For the vorticity $\omega = b \tau$, $b > 0$, the interlacing pairs 
  of functions 
  ${\cal R}_{2k}^{(+)}$ (dashed lines) and ${\cal R}_{2k+1}^{(+)}$ (bold lines),
  and ${\cal R}_{2k}^{(-)}$ (dashed lines) and ${\cal R}_{2k+1}^{(-)}$ (bold lines), 
  $k = 0,1,\dots$, are plotted for $b=1/2$ (tips are closer to the vertical axis) 
  and $b=2$.}
\end{figure}

\subsection{Linear positive vorticity}

Let $\omega = b \tau$, where $b$ is a positive constant. Then $\Omega (\tau) = b
\tau^2 / 2$, and so $s_0 = \sqrt b$. According to formulae (4.4), (4.5) and (4.8) in
\cite{KK}, the following two sequences
\begin{equation}
h_j^{(+)} (s) = \frac{(-1)^j}{\sqrt b} \arcsin \frac{\sqrt b}{s} + j
\frac{\pi}{\sqrt b} \quad {\rm and} \quad h_j^{(-)} (s) = h_j^{(+)} (s) +
\frac{\pi}{\sqrt b} , \quad j=0,1,\dots , \label{N2_t}
\end{equation}
give the depths of flows with horizontal free surfaces. Note that if $j$ is even
(odd), then both functions (\ref{N2_t}) are strictly increasing (decreasing,
respectively). Thus, the left-hand-side terms in equation (\ref{ss2}) are as follows
(see Fig.~2, where the corresponding graphs are plotted for $b=1/2$ and $b=2$;
reproduced from \cite{KK}):
\begin{equation}
{\cal R}_j^{(\pm)} (s) = \frac{1}{3} \left[ s^2 - b + \frac{2 (-1)^j}{\sqrt b}
\arcsin \frac{\sqrt b}{s} + \frac{2 \pi}{\sqrt b} \left( j + \frac{1}{2} \mp
\frac{1}{2} \right) \right] , \quad j=0,1,\dots . \label{N2+1_t}
\end{equation}
In particular, we have that
\begin{eqnarray}
&& {\cal R}_0^{(+)} (s) = \frac{1}{3} \left( s^2 - b + \frac{2}{\sqrt b} \arcsin
\frac{\sqrt b}{s} \right) \ \ \mbox{and} \nonumber \\ && {\cal R}_1^{(+)} (s) =
\frac{1}{3} \left( s^2 - b + \frac{2 \pi}{\sqrt b} - \frac{2}{\sqrt b} \arcsin
\frac{\sqrt b}{s} \right) . \label{feb14'}
\end{eqnarray}
(In each of the two families shown in Fig.~2, the graphs corresponding to these
functions are plotted as the lowest dashed and solid lines, respectively.) Hence
$r_0 = {\cal R}_{0}^{(+)} (s_0) = {\cal R}_{1}^{(+)} (s_0) = \frac{\pi}{3 \sqrt
b}$, whereas $r_c$\,---\,the only minimum of ${\cal R}_0^{(+)} (s)$\,---\,is
attained at
\begin{equation}
s_c = \sqrt{ \frac{b}{2} + \sqrt{\left( \frac{b}{2} \right)^2 + 1} } \, .
\label{s_c}
\end{equation}
Furthermore, both relations (\ref{R_s-}) remain true in the present case together
with the conclusions drawn from them about the pattern formed by the graphs of
${\cal R}_{0}^{(+)}$ and ${\cal R}_{1}^{(+)}$. Moreover, this pattern is repeated
infinitely many times; indeed, formulae (\ref{N2+1_t}) imply that it is shifted by
$2 \pi j / (3 \sqrt b)$, $j = 1,2,\dots$, along the positive $r$-axis (see Fig.~2).

The first component of a stream solution has the form
\begin{equation}
U (Y; s) = \pm \frac{s}{\sqrt b} \sin \sqrt b Y \, . \label{U_b_t}
\end{equation}
Indeed, this function solves the Cauchy problem that consists of the first and
second relations (\ref{ss1}). The sign in (\ref{U_b_t}) coincide with that in the
superscript of the corresponding $h^{(\pm)}_j (s)$, which is the second component of
the same stream solution. Infinitely many options of it are given by formulae
(\ref{N2_t}); however, only a finite number of them defines steady flows of constant
depth for a particular value of $r$ greater than $r_c$. The number depends on how
many roots has the corresponding set of equations (\ref{ss2}) for a chosen $r$. Note
that the number of roots increases with $r$.

\vspace{2mm}

Prior to investigating which flows support Stokes waves we turn to the sequence of
dispersion equations which exists in the present case. Indeed, the left-hand side of
(\ref{dispers}) involves $\gamma$ that solves the following problem:
\begin{equation}
- \gamma'' + ( \tau^2 - b ) \, \gamma = 0 , \quad \gamma (0) = 0 , \ \ \gamma (h) =
1 , \label{gam_b}
\end{equation}
where $h = h^{(\pm)}_j (s)$, $j = 1,2,\dots$. In each of these values, $s$ must be
set equal to a root of equation (\ref{ss2}) with ${\cal R}_j^{(\pm)} (s)$ in the
left-hand side. Applying the remark that follows problem (\ref{gamma}), we conclude
that (\ref{gam_b}) has a solution for all $\tau$ such that
\begin{equation}
\tau^2 \neq b - \left( \frac{\pi k}{h} \right)^2 , \quad k = 1,2,\dots .
\label{spec_b}
\end{equation}

This condition is obviously fulfilled when $\tau^2 \geq b$, in which case $\gamma
(Y, \tau)$ is equal to
\begin{eqnarray*}
&& \!\!\!\!\!\!\!\!\!\!\!\! \bullet \quad \quad \quad Y / h \quad \quad \quad \ \
\mbox{when} \ \tau^2 = b ; \\ && \!\!\!\!\!\!\!\!\!\!\!\! \bullet \quad \frac{\sinh
\sqrt{\tau^2 - b} \, Y} {\sinh \sqrt{\tau^2 - b} \, h} \quad \mbox{when} \ \tau^2 >
b ,
\end{eqnarray*}
and so $\gamma$ (and $\sigma$ as well) is a continuous function of $\tau$ when
$\tau^2 \geq b$. Using formula (\ref{U_b_t}) with $h = h^{(\pm)}_j (s)$ and the
second of the bulleted expressions in formula (\ref{dispers}), we get that the
sequence of dispersion equations for $\tau^2 \geq b$ has the form:
\begin{eqnarray}
s^2 \sqrt{\tau^2 - b} \, \cos^2 \left( \sqrt b \, h^{(\pm)}_j (s) \right) \coth
\left( \sqrt{\tau^2 - b} \, h^{(\pm)}_j (s) \right) \nonumber \\ - 1 \pm b s \cos
\left( \sqrt b \, h^{(\pm)}_j (s) \right) = 0 , \ \ j = 0,1,\dots , \label{d_b>}
\end{eqnarray}
and the left-hand side must be understood as the corresponding limit for $\tau^2 =
b$.

If $\tau^2 < b$ and condition (\ref{spec_b}) is fulfilled, then
\[ \gamma (Y, \tau) = \frac{\sin \sqrt{b - \tau^2} \, Y}{\sin \sqrt{b - \tau^2} \, h}
\, . 
\]
Substituting (\ref{U_b_t}) and this expression into formula (\ref{dispers}), we get
the sequence of dispersion equations for this case:
\begin{eqnarray}
s^2 \sqrt{b - \tau^2} \, \cos \left( \sqrt b \, h^{(\pm)}_j (s) \right) \cot \left(
\sqrt{b - \tau^2} \, h^{(\pm)}_j (s) \right) \nonumber \\ - 1 \pm b s \cos \left(
\sqrt b \, h^{(\pm)}_j (s) \right) = 0 , \ \ j = 0,1,\dots \label{d_b<}
\end{eqnarray}
The limit form of (\ref{d_b<}) as $\tau^2 \to b - 0$ coincides with that of
(\ref{d_b>}) as $\tau^2 \to b + 0$. It is clear that the derivatives of the
left-hand sides of (\ref{d_b>}) and (\ref{d_b<}) have definite signs on the
intervals of continuity, and so if these equations have roots, then they are simple.
In order to investigate equations (\ref{d_b>}) and (\ref{d_b<}) we apply
Proposition~3.3.

\vspace{2mm}

Let us consider several particular cases. Taking into account formula
(\ref{spec_b}), we begin with flows such that $\sqrt b \, h^{(\pm)}_j (s) < \pi$.
Formulae (\ref{N2_t}) show that only $h^{(+)}_0 (s)$ and $h^{(+)}_1 (s)$ satisfy
this inequality. Moreover, Proposition~3.3~(i) guarantees that for each of these
depths the pair of equations (\ref{d_b>}) and (\ref{d_b<}) has one and only one
positive root provided condition (\ref{condRh}) is fulfilled for either of the
following pairs:
\[ {\cal R} = {\cal R}^{(+)}_0 , \ \ h = h^{(+)}_0 ; \quad 
{\cal R} = {\cal R}^{(+)}_1 , \ \ h = h^{(+)}_1 .
\]
In the first of these cases, we have the depth $h^{(+)}_0 (s)$, where $s \in (s_0,
s_c)$ is obtained from equation (\ref{ss2}) with ${\cal R}_0^{(+)}$ and $r \in (r_c,
r_0)$. Any such flow is unidirectional because of the upper sign in formula
(\ref{U_b_t}) which gives that
\[ U' (Y; s) = s \cos \sqrt b Y > 0 \quad \mbox{for all} \ Y \in \big[ 0, \, 
h^{(+)}_0 (s) \big] .
\]
In the second of the above cases, we have the depth $h^{(+)}_1 (s)$, where $s > s_0$
is obtained from equation (\ref{ss2}) with ${\cal R}_1^{(+)}$ and an arbitrary $r >
r_0$. All such flows have a near-surface counter-current because $U' (Y; s) = s \cos
\sqrt b Y$ changes sign only once on $\big( 0, \, h^{(+)}_1 (s) \big)$.

The results obtained in \cite{KK}, section~6.3, yield that inequality (\ref{condRh})
always holds for the functions ${\cal R}^{(+)}_1$ and $h^{(+)}_1$, whereas
(\ref{condRh}) holds for ${\cal R}^{(+)}_0$ and $h^{(+)}_0$ only when $r \in (r_c,
r_0)$. [We recall that $r_0 = \pi / 3 \sqrt b$, whereas $r_c = {\cal R}_0^{(+)}
(s_c)$ and $s_c$ is given by formula (\ref{s_c}).] Thus, {\it there exists the
unique bifurcation wavelength for each of the shear flows whose depths are
$h^{(+)}_0 (s)$ with $s \in (s_0, s_c)$ and $h^{(+)}_1 (s)$ with $s > s_0$.
According to Theorem~1.2, a family of Stokes waves with fixed $r$ bifurcates from
every such wavelength.} 

Besides, Proposition~3.5 is applicable when ${\cal R} = {\cal R}^{(+)}_0$ and $h =
h^{(+)}_0$. Therefore, {\it if $\sigma$ is defined by the stream solution
corresponding to the root $s > s_c$ of ${\cal R}_0^{(+)} (s) = r$ with an arbitrary
$r > r_c$, then the dispersion equations $(\ref{d_b>})$ and $(\ref{d_b<})$ have no
positive solutions.}

\vspace{2mm}

Now let $\sqrt b \, h^{(\pm)}_j (s) \in (\pi, 2 \pi)$, in which case formulae
(\ref{N2_t}) show that only $h^{(-)}_0 (s)$ and $h^{(-)}_1 (s)$ are the appropriate
values of depth, whereas $s$ must satisfy ${\cal R}_0^{(-)} (s) = r$ and ${\cal
R}_1^{(-)} (s) = r$, respectively. It is clear that the second of these equations
has only one solution for any $r > r_0 + \frac{2 \pi}{3 \sqrt b}$ and the range of
solutions is $(s_0 , +\infty)$. Furthermore, the equation ${\cal R}_0^{(-)} (s) = r$
has only one solution belonging to $(s_c , +\infty)$ for any $r > r_c + \frac{2
\pi}{3 \sqrt b}$. Moreover, for every $r \in \big( r_c + \frac{2 \pi}{3 \sqrt b} ,
r_0 + \frac{2 \pi}{3 \sqrt b} \big)$ the latter equation has another solution in
$(s_0 , s_c)$. Thus, the pair of equations ${\cal R}_0^{(-)} (s) = r$ and ${\cal
R}_1^{(-)} (s) = r$ has a pair of solutions for all $r > r_c + \frac{2 \pi}{3 \sqrt
b}$ except for $r = r_0 + \frac{2 \pi}{3 \sqrt b}$, when only the first of these
equations has a solution. All flows, whose constant depths are equal to $h^{(-)}_0
(s)$ and to $h^{(-)}_1 (s)$, have near-bottom counter-currents. In the first case,
there is no other counter-current, but the near-surface counter-current is also
present in the second case.

Substituting $h^{(-)}_0 (s)$ and $h^{(-)}_1 (s)$ considered in the previous
paragraph into equations (\ref{d_b>}) and (\ref{d_b<}), we see that
Proposition~3.3~(ii) is applicable because formula (\ref{spec_b}) yields that
\[ \tau^2_* = b - \left( \pi / h \right)^2 \ \ \mbox{with} \ h = h^{(-)}_k (s), 
\quad k=0,1,
\] 
is the only Dirichlet eigenvalue of the operator $\D^2 / \D Y^2 + \omega' (U)$
considered on $(0,h)$. Then the cited proposition implies that equations
(\ref{d_b>}) and (\ref{d_b<}) with $h^{(-)}_k (s)$, $k=0,1$, have exactly one
positive root $\tau_+^{(k)} \in \left( \tau_* , +\infty \right)$. On the other hand,
if condition (\ref{condRh}) is fulfilled for the following pairs
\[ {\cal R} = {\cal R}^{(-)}_0 , \ \ h = h^{(-)}_0 , \quad \mbox{and} \quad 
{\cal R} = {\cal R}^{(-)}_1 , \ \ h = h^{(-)}_1 ,
\]
then Proposition~3.3~(i) is applicable. From the results obtained in \cite{KK},
section~6.3, it follows that (\ref{condRh}) is always fulfilled for the second pair
of functions, whereas for the first pair it holds only when $s \in (s_0 , s_c)$,
that is, when $r \in \big( r_c + 2 \pi / 3 \sqrt b , r_0 + 2 \pi / 3 \sqrt b \big)$
in equation (\ref{ss2}) with ${\cal R}^{(-)}_0$ in the left-hand side. Then
Proposition~3.3~(i) guarantees that equations (\ref{d_b>}) and (\ref{d_b<}) with
$h^{(-)}_k (s)$, $k=0,1$, have one and only one positive root $\tau_-^{(k)} \in
\left( 0, \tau_* \right)$. Thus, {\it there exist two bifurcation wavelengths for
each of the shear flows whose depths are $h^{(-)}_0 (s)$ with $s \in (s_0, s_c)$ and
$h^{(-)}_1 (s)$ with $s > s_0$. Moreover, there exists the bifurcation wavelength
for the shear flows whose depth is $h^{(-)}_0 (s)$ with $s > s_c$. According to
Theorem~1.2, families of Stokes waves with fixed values of $r$ bifurcate from all
these wavelength.}

\vspace{2mm}

If $\sqrt b \, h^{(\pm)}_j (s) \in (2 \pi, 3 \pi)$, then formulae (\ref{N2_t}) show
that only $h^{(+)}_2 (s)$ and $h^{(+)}_3 (s)$ satisfy this condition provided $s$ is
a root of ${\cal R}_2^{(+)} (s) = r$ and ${\cal R}_3^{(+)} (s) = r$, respectively.
The same analysis shows that these equations have the same roots as in the previous
case provided $2 \pi / 3 \sqrt b$ is added to $r$ that stands in the right-hand side
of ${\cal R}_0^{(-)} (s) = r$ and ${\cal R}_1^{(-)} (s) = r$, respectively. But now,
formula (\ref{spec_b}) yields the existence of two Dirichlet eigenvalues for the
operator $\D^2 / \D Y^2 + \omega' (U)$ considered on $(0,h)$:
\[ \tau^2_{*1} = b - \left( \pi / h \right)^2 \ \ \mbox{and} \ \ \tau^2_{*2} = b - 
\left( 2 \pi / h \right)^2 \ \ \mbox{with} \ h = h^{(+)}_k (s) , \quad k=2,3.
\]
The above considerations should be modified in view that more roots of equations
(\ref{d_b>}) and (\ref{d_b<}) exist. The required amendments immediately follow from
Proposition~3.3~(ii), and are left to the reader. What is important is that there is
no near-bottom counter-current and the number of counter-currents is different.

\vspace{2mm}

Now we are in a position to formulate the following conclusions about Stokes waves
on flows with positive linear vorticity.

\vspace{2mm}

\noindent {\bf Proposition 6.3.} {\it First, if $r > r_c$, then every root of each
equation ${\cal R}_j^{(\pm)} (s) = r$ defines a stream solution from which a family
of small-amplitude Stokes waves bifurcate with the wavelength defined by the
corresponding dispersion equation. The only exceptions are roots equal to $s_0$ and
those that are greater than $s_c$ and satisfy the equation ${\cal R}_0^{(+)} (s) =
r$.

Second, the number of stream solutions from which small-amplitude Stokes waves
bifurcate increases as $r$ grows. Indeed, for larger values of $r$ more equations
${\cal R}_j^{(\pm)} (s) = r$ provide roots that can be used in the dispersion
equations $(\ref{d_b>})$ and $(\ref{d_b<})$. On the other hand, the latter equations
have more roots defining bifurcation wavelengths because the number of the
Dirichlet eigenvalues of $\D^2 / \D Y^2 + \omega' (U)$ increases with $r$.}

\vspace{2mm}

The next assertion is analogous to Proposition 6.1.

\vspace{2mm}

\noindent {\bf Proposition 6.4.} {\it For every $s \in (s_0, s_c)$ there exists a
brunch of flows with small-amp\-litude Stokes waves on their free surfaces; for all
these flows $r = {\cal R}_0^{(+)} (s)$ $($see the first formula $(\ref{feb14'})$$)$.
This brunch bifurcates from the horizontal shear flow whose depth $h_0^{(+)} (s)$ is
given by the first formula $(\ref{N2_t})$ with the subscript zero.

For every $s \in (s_0, +\infty)$ analogous brunches bifurcate from the horizontal
shear flows whose depths are equal to $h_j^{(\pm)} (s)$, where $j = 0,1,\dots$ for
the superscript $(-)$ and $j = 1,2,\dots$ for the superscript $(+)$ $[$see formulae
$(\ref{N2_t})$$]$. The value of $r$ along each of these brunches is equal to ${\cal
R}_j^{(\pm)} (s)$ given by formula $(\ref{N2+1_t})$.}

\subsection{On violation of assumption (III) for linear vorticity}

Let the dispersion equation $\sigma_* (\tau; s) = 0$ [see (\ref{dispers_*})] with $s
= s_*$ have a root that defines the wavelength $\Lambda_0$ (see the first paragraph
of section~5.1). Then $s_*$ serves as the bifurcation point for the operator
equation (\ref{dec11}) and assumption (III), essential for applying Theorem~4.1 to
the latter equation, is expressed in terms of $s_*$ as follows: $\dot{\sigma}_*
(\tau; s_*)$ must be non-zero. In this section, we show that assumption (III) is
violated for some $s$ provided the depth function used in the definition of
$\sigma_*$ is one of those given by formulae (\ref{N2_t}).

In particular, let us show that $\dot{\sigma}_* (\tau (s); s)$ and $\sigma_* (\tau
(s); s)$ vanish simultaneously for some $s$ provided
\begin{equation}
h = h_0^{(-)} (s) = \frac{\pi}{\sqrt b} + \frac{1}{\sqrt{b}} \arcsin
\frac{\sqrt{b}}{s} , \quad s > \sqrt{b} , \label{22jan}
\end{equation}
[see the second formula (\ref{N2_t})]. It is clear that for this $h$ we have
\[ \sigma_* (\tau; s) \to +\infty \ \ \mbox{as} \ \tau \to +\infty \ \ \mbox{and}
\ \ \sigma_* (\tau; s) \to -\infty \ \ \mbox{as} \ \tau \to b-(\pi/h)^2 .
\]
Therefore, the equation $\sigma_* (\tau; s) = 0$ has the root $\tau (s) \in
(b-(\pi/h)^2, +\infty)$ with this $h$. This root is well-defined because
$\sigma_{*\tau} (\tau; s) > 0$ when $\sigma_* (\tau; s) = 0$ (this follows in the
same way as in the proof of Lemma~1.1). Moreover, $\tau (s)$ is a smooth function,
and so, differentiating the dispersion equation defining it with respect to $s$, we
obtain
\[ \dot{\tau} (s) \, \sigma_{*\tau} (\tau; s) + \dot{\sigma}_* (\tau; s) = 0 .
\]
Hence if $\dot{\tau} (s)$ vanishes for some $s > s_0 = \sqrt{b}$, then assumption
(III) is violated for $\tau (s)$.

Formula (\ref{22jan}) yields that $s\cos \left( \sqrt{b} \, h_0^{(-)} (s) \right) =
- \sqrt{s^2 - b}$, and so equations (\ref{d_b>}) and (\ref{d_b<}) take the form
\begin{equation}
(s^2 - b) \sqrt{\tau^2 (s) - b} \, \coth \left( \sqrt{\tau^2 (s) - b} \, h^{(-)}_0
(s) \right) - 1 + b \sqrt{s^2 - b} = 0  \label{d_b>0}
\end{equation}
and
\begin{equation}
(s^2 - b) \sqrt{b - \tau^2 (s)} \, \cot \left( \sqrt{b - \tau^2 (s)} \, h^{(-)}_0
(s) \right) - 1 + b \sqrt{s^2 - b} = 0 , \label{d_b<0}
\end{equation}
respectively. Let us consider what these equations yield about the behaviour of
$\tau (s)$ when $s$ is close to $\sqrt b = s_0$ and $s \to +\infty$.

Using properties of the hyperbolic tangent, one obtains from (\ref{d_b>0}) that
\begin{equation}
\tau (s) = \frac{1}{s^2-b} + O \left( \frac{1}{\sqrt{s^2-b}} \right) \quad \mbox{as}
\ s \to \sqrt b + 0 . \label{27jan}
\end{equation}
In the second limiting case, our aim is to use (\ref{d_b<0}) for demonstrating that
for all $b > 0$ the following asymptotic representation
\begin{equation}
\tau (s) = \frac{\sqrt{3b}}{2} + \frac{a}{s} \quad \mbox{holds as} \ s \to + \infty
. \label{22janu}
\end{equation}
If $a$ is a negative constant (and so the square root in the first term in
(\ref{d_b<0}) exists for all $s$), then (\ref{22janu}) implies that the graph of
$\tau (s)$ approaches the horizontal asymptote from below. Combining this and
(\ref{27jan}), we see that $\tau (s)$ attains minimum at some $s > \sqrt{b}$, and so
$\dot{\tau} (s)$ vanishes there and, may be, somewhere else.

In order to prove (\ref{22janu}) we divide (\ref{d_b<0}) by $(s^2 - b)$ and keeping
the leading terms get
\begin{equation}
\sqrt{b - \tau^2 (s)} \cot \left( \sqrt{b - \tau^2 (s)} \, h^{(-)}_0 (s) \right)
\sim - \frac{b}{s} \, . \label{24jan}
\end{equation}
For finding $a$, we substitute (\ref{22janu}) and obtain that
\[ \sqrt{b - \tau^2 (s)} \, h^{(-)}_0 (s) \sim \frac{\sqrt{b}}{2} \left[ 1 -
\frac{2a \sqrt{3b}}{bs} \right] \left[ \frac{\pi}{\sqrt{b}} + \frac{1}{s} \right] .
\]
Therefore, after simplification (\ref{24jan}) takes the form
\[ \frac{\sqrt{b}}{2} \cos \left( \frac{\pi}{2} + \frac{\sqrt{b}}{2s} - \frac{\pi
a \sqrt{3b}}{bs} \right) \sim - \frac{b}{s} \, ,
\]
which is equivalent to
\[ \frac{\sqrt{b}}{2} \left( \frac{\sqrt{b}}{2s} - \frac{\pi a\sqrt{3b}}{bs}
\right) \sim \frac{b}{s} \, .
\]
This gives $a = - \frac{b \sqrt{3}}{2 \pi} < 0$, thus completing the proof of the
asymptotic formula (\ref{22janu}) with $a$ guaranteeing that $\dot{\tau} (s)$
vanishes at some $s > \sqrt{b}$.

\section*{Appendix A. On scaling}

Let $\tilde x$ and $\tilde y$ be the dimensional horizontal and vertical
coordinates, respectively. The usual free-boundary problem describing steady water
waves with vorticity in terms of the dimensional stream function $\tilde \psi$ and
the free surface profile $\tilde \eta$ is as follows (cf. (2.2) in \cite{CS}):
\begin{eqnarray}
&& \tilde \psi_{\tilde x \tilde x} + \tilde \psi_{\tilde y \tilde y} + \upsilon
(\tilde \psi) = 0, \quad \mbox{for} \ 0 < \tilde y < \tilde \eta (\tilde x) , \
\tilde x \in \RR ; \label{1''} \\ && \tilde \psi (\tilde x, 0) = 0,\quad \tilde x
\in \RR ; \label{2''} \\ && \tilde \psi (\tilde x, \tilde \eta (\tilde x)) = Q ,
\quad \tilde x \in \RR ; \label{3''} \\ && \frac{1}{2} |\nabla_{\tilde x, \tilde y}
\, \tilde \psi (\tilde x, \tilde \eta (\tilde x))|^2 + g \, \tilde \eta (\tilde x) =
R , \quad \tilde x \in \RR . \label{4''}
\end{eqnarray}
Here $\upsilon$ is the given vorticity distribution; $(0, -g)$ is the constant
gravity vector; $Q$ and $R$ are given constants expressing the volume rate of flow
per unit span and the total head (Bernoulli's constant), respectively. Of course,
$g$ and $R$ are positive, whereas $Q$ is arbitrary, but we assume that $Q \neq 0$
because the case of zero rate of flow needs special treatment. Note that $Q$ cannot
be zero in the irrotational case (see \cite{KK6}).

In order to transform the above problem into the non-dimensional problem
(\ref{1})--(\ref{4}), we apply the following scaling:
\[ X = \frac{\tilde x}{(Q^2/g)^{1/3}}, \quad Y = \frac{\tilde y}{(Q^2/g)^{1/3}}, 
\quad \xi (X) = \frac{\tilde \eta (\tilde x)}{(Q^2/g)^{1/3}}, \quad \Psi (X,Y) =
\frac{\tilde \psi (x,y)}{Q} \, .
\]
Here $(Q^2/g)^{1/3}$ is the depth of the critical uniform stream in the irrotational
case (see \cite{Ben}).

It is clear that the boundary conditions (\ref{2''}) and (\ref{3''}) turn into
(\ref{2}) and (\ref{3}) after scaling. From (\ref{1''}) one gets equation (\ref{1})
for $\Psi$ with $\omega (\Psi) = (Q/g^2)^{1/3} \upsilon (Q \Psi)$. It is
straightforward to calculate that dividing (\ref{4''}) by $R_c = \frac{3}{2} (Q
g)^{2/3}$ one obtains (\ref{4}) with $r = R/R_c$. We recall that $R_c$ is the value
of Bernoulli's constant for the critical uniform stream in the irrotational case
(see \cite{Ben}). Finally, note that the scaling length used above gives the
dimensional bifurcation wavelength equal to $(Q^2/g)^{1/3} \Lambda_0$, where the
non-dimensional value $\Lambda_0$ is defined on the basis of the dispersion equation
(see the next paragraph after Lemma~1.1).

\section*{Appendix B. Basic properties of stream solutions}

Integrating the first equation (\ref{ss1}) for $U (Y; s)$, one obtains
\begin{equation}
(U')^2 + 2 \, \Omega (U) = s^2 . \label{final}
\end{equation}
Moreover, we get that if $h (s) > 0$ is found for some $s$ so that
\begin{equation}
U (h (s); s) = 1 \quad {\rm and} \quad s^2 - 2 \, \Omega (U (h (s); s) ) + 2 h (s) =
3 r \label{ber_tr}
\end{equation}
hold simultaneously, then the pair $(U (Y; s),\, h (s))$ is a stream solution (note
that in the second relation (\ref{ber_tr}), the expression from (\ref{final}) is
substituted). These solutions exist only for $r \geq r_c$, where $r_c$ is defined by
formula (\ref{ss3}), and the set of stream solutions corresponding to a particular
$r$ can be obtained by virtue of the following procedure (see details in
\cite{KK}).

\vspace{2mm}

One takes some $s \geq s_0$ [see (\ref{s_0})] and finds $U (Y; s)$ from the Cauchy
problem introduced in section 1.3. In fact, $s$ depends on $r$ and to get $s$ for a
given $r$ one has to use equation (\ref{ss2}) (note that several values of $s$ might
correspond to some particular $r$). Then one obtains the values of depth for which
purpose the first relation (\ref{ber_tr}) serves. Let us outline the corresponding
scheme involving two auxiliary quantities which we denote $\tau_\pm (s)$ and $y_\pm
(s)$ (see \cite{KK}, section~3, for their properties).

If the equation $2 \, \Omega (\tau) = s^2$ with $s \geq s_0$ has finite positive and
negative roots, then by $\tau_+ (s)$ [$\tau_- (s)$] we denote the smallest positive
root [the largest negative root, respectively]. If there is no finite positive
[negative] root, then we put $\tau_+ (s) = +\infty$ [$\tau_- (s) = -\infty$,
respectively]. Notice that $\tau_+ (s) \geq 1$ if and only if $s \geq s_0$. Now we
set
\[ y_\pm (s) = \int_0^{\tau_\pm (s)} \frac{\D \tau}{\sqrt{s^2 - 2\Omega (\tau)}} 
\quad \mbox{for} \ s \geq s_0 .
\]
The introduced quantities are such that $\left( y_- (s) , \, y_+ (s) \right)$ is the
maximal interval, where $U (Y; s)$ increases strictly monotonically, and $\tau_+
(s)$ [$\tau_- (s)$] is the supremum [infimum, respectively] of $U (Y; s)$ on this
interval. More precisely, $U (Y; s)$ has the following properties.

\vspace{2mm}

\noindent $\bullet$ If $y_+ (s) = +\infty$ and $y_- (s) = -\infty$, then $U (Y; s)$
increases strictly monotonically for all $Y \in \RR$.

\noindent $\bullet$ If $y_- (s) = -\infty$ and $y_+ (s) < +\infty$, then $U (Y; s)$
is bimonotonic and attains its maximum $\tau_+ (s)$ at $Y = y_+ (s)$.

\noindent $\bullet$ If $y_- (s) > -\infty$ and $y_+ (s) = +\infty$, then $U (Y; s)$
is bimonotonic and attains its minimum $\tau_- (s)$ at $Y = y_- (s)$.

\noindent $\bullet$ If both $y_+ (s)$ and $y_- (s)$ are finite, then $U (Y; s)$ is
harmonic-like; it attains one of its minima at $Y = y_- (s)$ and one of its maxima
at $Y = y_+ (s)$. Moreover, $U (Y; s)$ increases strictly monotonically from $\tau_-
(s)$ to $\tau_+ (s)$ on $[y_- (s) , \, y_+ (s)]$.

\vspace{2mm}

Furthermore, it is proved in \cite{KK}, section~4.2, that for $s \geq s_0$
\begin{equation} 
h_0 (s) = \int_0^1 \frac{\D \tau}{\sqrt{s^2 - 2\Omega (\tau)}} \label{hs}
\end{equation}
is the smallest positive value of the depth such that $U (h_0 (s); s) = 1$. By
$h_j^{(+)} (s)$, $j = 0, 1,\dots$, we denote the whole sequence (possibly finite) of
depth values such that $U \left( h_j^{(+)} (s); s \right) = 1$ for a given $s$, and
we have
\begin{eqnarray}
&& h_{2k}^{(+)} (s) = h_0 (s) + 2 k \left[ y_+ (s) - y_- (s) \right] , \quad j = 2 k ;
\label{h2k} \\ && \!\!\!\!\!\!\!\! h_{2k+1}^{(+)} (s) = h_0 (s) + 2 \left[ y_+ (s) - h
(s) \right] + 2 k \left[ y_+ (s) - y_- (s) \right] , \quad j = 2 k+1 . \label{h2k+1}
\end{eqnarray}
Here, $k = 0, 1,\dots$, and so $h_0 (s)$ [see formula (\ref{hs})] is included into
(\ref{h2k}), where it appears as $h_{0}^{(+)} (s)$. If both $y_+ (s)$ and $y_- (s)$
are finite, then formulae (\ref{h2k}) and (\ref{h2k+1}) give finite values for all
$k = 0, 1,\dots$. Otherwise the first of them gives a finite value only for $k = 0$.
Formula (\ref{h2k+1}) also gives a finite value for $k = 0$ provided $y_+ (s)$ is
finite. Note that both formulae coincide when $y_+ (s) = h (s)$ that is equivalent
to the equality $\tau_+ (s) = 1$.

For $s > s_0$, there exists the following sequence of solutions
\begin{equation}
h_j^{(-)} (s) = h_j^{(+)} (s) - 2 y_- (s) , \quad j = 0,1,\dots , \label{hj-}
\end{equation}
of the equation $U (h (s); -s) = 1$, and so this sequence also gives flow depths.
The number of finite elements in the sequence (\ref{hj-}) depends on whether $y_+
(s)$ and $y_- (s)$ are finite or not. It is clear that the value $h_j^{(-)} (s)$ is
infinite when either $h_j^{(+)} (s) = +\infty$ or $y_- (s) = -\infty$, whereas
$h_j^{(-)} (s)$ is finite otherwise.

The subscript (superscript) at $h_j^{(\pm)} (s)$ points at the number of
counter-currents (the direction of flow in the near-bottom layer, respectively) in
the following way. If the depth is equal to $h_j^{(+)}$, then the number of layers
with alternating directions of flow is equal to $j+1$, whereas the direction of flow
in the near-bottom layer coincides with the direction of the so-called critical
unidirectional flow, which always exists, corresponds to $r = r_c$ and has the
minimum depth among all possible flows. If the depth is equal to $h_j^{(-)}$, then
there are $j+2$ layers with alternating directions of flow and the near-bottom layer
has the direction opposite to that of the critical flow.

\vspace{2mm}

Finally, it must be mentioned that if
\begin{equation}
r_0 = \lim_{s \to s_0 + 0} {\cal R}_0^{(+)} (s) \label{ss4}
\end{equation}
is finite, then among the flows corresponding to every $r > r_0$ there exists at
least one flow that has a counter-current. These flows are defined by stream
solutions obtained above and flows with several counter-currents can exist among
them.

\vspace{6mm}

\noindent {\bf Acknowledgements.} The authors are grateful to the anonymous referee;
his/her comments were helpful for improving the presentation. V.~K. was supported by
the Swedish Research Council~(VR). N.~K. acknowledges the financial support from the
Link\"oping University.

\end{document}